%% file: main.tex
\documentclass[sigconf, titlepage]{acmart}
\usepackage{fancyhdr}
\AtBeginDocument{%
  \providecommand\BibTeX{{%
    \normalfont B\kern-0.5em{\scshape i\kern-0.25em b}\kern-0.8em\TeX}}}

\usepackage{algorithmic}
\usepackage{graphicx}
\usepackage{textcomp}
\usepackage{xcolor}
\usepackage{xspace}
\usepackage{subfigure}
\usepackage{multirow}
\usepackage{url}
\usepackage{bbding}
\usepackage{pifont} 
\usepackage{array}
\usepackage{diagbox}
\usepackage{float}
\usepackage[most]{tcolorbox}
\usepackage[export]{adjustbox}
\usepackage{caption}
\usepackage{hyperref}
\usepackage{booktabs,caption}
\usepackage[flushleft]{threeparttable}
\usetikzlibrary{automata,positioning}
\renewcommand\footnotetextcopyrightpermission[1]{}

\newcommand{\vm} {\textsf{VSMask}\xspace}

\usepackage{filecontents}    
\newcommand{\rev}[1]{{\color{black} #1}}
\newcommand{\newrev}[1]{{\color{black} #1}}
\begin{document}

\title{VSMask: Defending Against Voice Synthesis Attack via Real-Time Predictive Perturbation}
\author{Yuanda Wang}  
                                  % ...

\affiliation{
  \institution{Michigan State University}
  \city{East Lansing}
  \state{Michigan}
  \country{USA}  
  }
\email{wangy208@msu.edu} 
\author{Hanqing Guo}

\affiliation{
  \institution{Michigan State University}
  \city{East Lansing}
  \state{Michigan}
  \country{USA}
  }
\email{guohanqi@msu.edu}
\author{Guangjing Wang}  
                                  % ...

\affiliation{
  \institution{Michigan State University}
  \city{East Lansing}
  \state{Michigan}
  \country{USA}  
  }
  \email{wanggu22@msu.edu} 
\author{Bocheng Chen}  
                                  % ...

\affiliation{
  \institution{Michigan State University}
  \city{East Lansing}
  \state{Michigan}
  \country{USA}  
  }
  \email{chenboc1@msu.edu} 
\author{Qiben Yan}  
                                  % ...

\affiliation{
  \institution{Michigan State University}
  \city{East Lansing}
  \state{Michigan}
  \country{USA}  
  }
  \email{qyan@msu.edu} 
\renewcommand{\shortauthors}{Yuanda Wang, Hanqing Guo, Guangjing Wang, Bocheng Chen, \& Qiben Yan}

\vspace{-5pt}
\begin{abstract}
Deep learning based voice synthesis technology generates artificial human-like speeches, which has been used in deepfakes or identity theft attacks. Existing defense mechanisms inject subtle adversarial perturbations into the raw speech audios to mislead the voice synthesis models. However, optimizing the adversarial perturbation not only consumes substantial computation time, but it also requires the availability of entire speech. Therefore, they are not suitable for protecting live speech streams, such as voice messages or online meetings. In this paper, we propose \vm, a real-time protection mechanism against voice synthesis attacks.
Different from offline protection schemes, \vm leverages a predictive neural network to forecast the most effective perturbation for the upcoming streaming speech.
\vm introduces a universal perturbation tailored for arbitrary speech input to shield a real-time speech in its entirety.
To minimize the audio distortion within the protected speech, we implement a weight-based perturbation constraint to reduce the perceptibility of the added perturbation. We comprehensively evaluate \vm protection performance under different scenarios. 
The experimental results indicate that \vm can effectively defend against 3 popular voice synthesis models. 
None of the synthetic voice could deceive the speaker verification models or human ears with \vm protection.
In a physical world experiment, we demonstrate that \vm successfully safeguards the real-time speech by injecting the perturbation over the air.
\end{abstract}

%
%


\pagestyle{plain}
\maketitle
\begin{figure}[t]
    \setlength{\abovecaptionskip}{-0pt} 
    \setlength{\belowcaptionskip}{-20pt} 
    \centering
    \subfigure[Synthetic voice commands are used to bypass the speaker verification process and illegally control the voice assistants.]{\includegraphics[width=7cm]{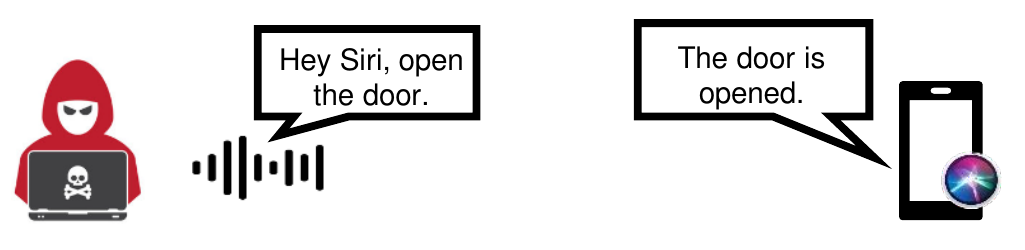}\label{fig:voice assistants}}
    \subfigure[Synthetic voice is used to hack into private accounts. ]{\includegraphics[width=7cm]{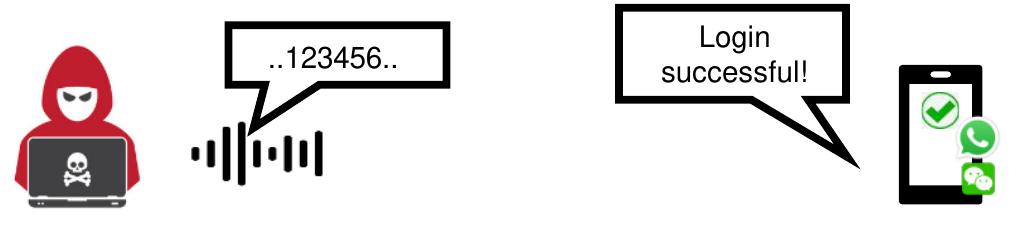}\label{fig: apps}}
    \subfigure[The adversary can steal the victim's identity using voice synthesis.]{\includegraphics[width=7cm]{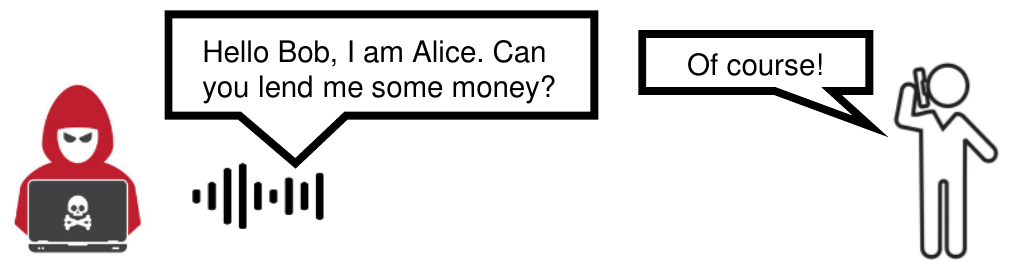}\label{fig: ghostcall}}
    \caption{Voice synthesis attack deceives both ASV system and human ears in real-world scenarios.}
    \label{fig: voice synthesis scenario}
\end{figure}
\input{document/1_introduction}

\input{document/2_background}
\input{document/3_model}
\input{document/4_evaluation}
\input{document/5_related_work}
\input{document/6_conclusion}
\bibliographystyle{IEEEtran}
\bibliography{reference}
\appendix
\section*{Appendix}
\section{Predictive Model}
\label{model_appendix}
\rev{According to \vm system design in Section 4, the input and output of the predictive model are supposed to be 2-D vectors with the same dimension and different lengths. Inspired by the model architecture in VoiceCamo~\cite{chiquier2021real}, we apply a similar down-sampling and up-sampling architecture in the \vm predictive model. To ensure the best prediction performance, we test different kernel sizes, strides, and layer numbers. We start with kernel size (2, 2) and determine that kernel size (3, 3) achieves the best trade-off between the protection performance and model complexity. Then, we carefully adjust the strides and layer numbers to correct the final output vector dimension. Finally, we use a tanh activation layer to normalize the predictive perturbation. The detailed parameter configuration is listed in Table~\ref{Tab: architecture}.
}
The architectural parameters vary according to the different preprocessing approaches in different voice synthesis models.
\begin{table}[htbp]
\centering
\caption{The detailed network parameters of \vm real-time predictive model when targeting AdaIN-VC model.}
\label{Tab: architecture}
\begin{tabular}{cccc} 
\toprule
Type                            & Input Size & Kernel Size          & Stride                  \\ 
\midrule
\multirow{7}{*}{Down-sampling~} & 1×512×100  & 3×3                  & (1,2)                   \\ 
\cmidrule{2-4}
                                & 32×512×50  & \multirow{6}{*}{3×3} & \multirow{6}{*}{(2,2)}  \\
                                & 128×512×25 &                      &                         \\
                                & 256×256×13 &                      &                         \\
                                & 256×128×7  &                      &                         \\
                                & 512×64×4   &                      &                         \\
                                & 512×32×2   &                      &                         \\ 
\midrule
\multirow{5}{*}{Up-sampling}    & 512×16×1   & \multirow{5}{*}{3×3} & \multirow{5}{*}{(2,2)}  \\
                                & 256×32×2   &                      &                         \\
                                & 128×64×4   &                      &                         \\
                                & 64×128×8   &                      &                         \\
                                & 32×256×16  &                      &                         \\ 
\midrule
tanh                            & 1×512×32   & N/A                  & N/A                     \\
\bottomrule
\end{tabular}
\end{table}

\section{Impact of Weights on Audio Quality}
\label{weight_appendix}
We carefully select the weight in Problem (\ref{weight}) for different frequency bands to ensure they have the same protection performance.
In this section, we investigate how the weight-based amplitude constraint mitigates the noise perceptibility in protected audios.

In human study, we also ask the volunteers to compare the clear speech and protected speech without weight constraint.
Fig.~\ref{fig: weight_hs} shows the perceptibility difference between protected audio with and without weighted constraint.
When we implement weight-based constraint, the perturbation in medium frequency is weakened.
In this way, human ears are less sensitive to the noise in the speech samples.
Over 60\% answers consider the noise is negligible or imperceptible, which means that \vm perturbation does not affect the utterance and voice in the speech at all.
There are also 30\% answers present they can tolerate the noise in the speech despite they are conscious of the perturbation.

In comparison, if we remove the weight mechanism, it becomes easier to sense the perturbation.
Less than 50\% answers demonstrate they can neglect the perturbation in the speech, while more than 20\% of results indicate that there is conspicuous noise in the audio.
A few responses point out that the noise is too strong which affects the utterances and voice in the speech.
Therefore, the human study results demonstrate that \vm's weighted constraint can significantly reduce the audio distortion in the protected audios.
\begin{figure}[t]
\centering
\label{fig: weight_hs}
\includegraphics[width =0.40\textwidth]{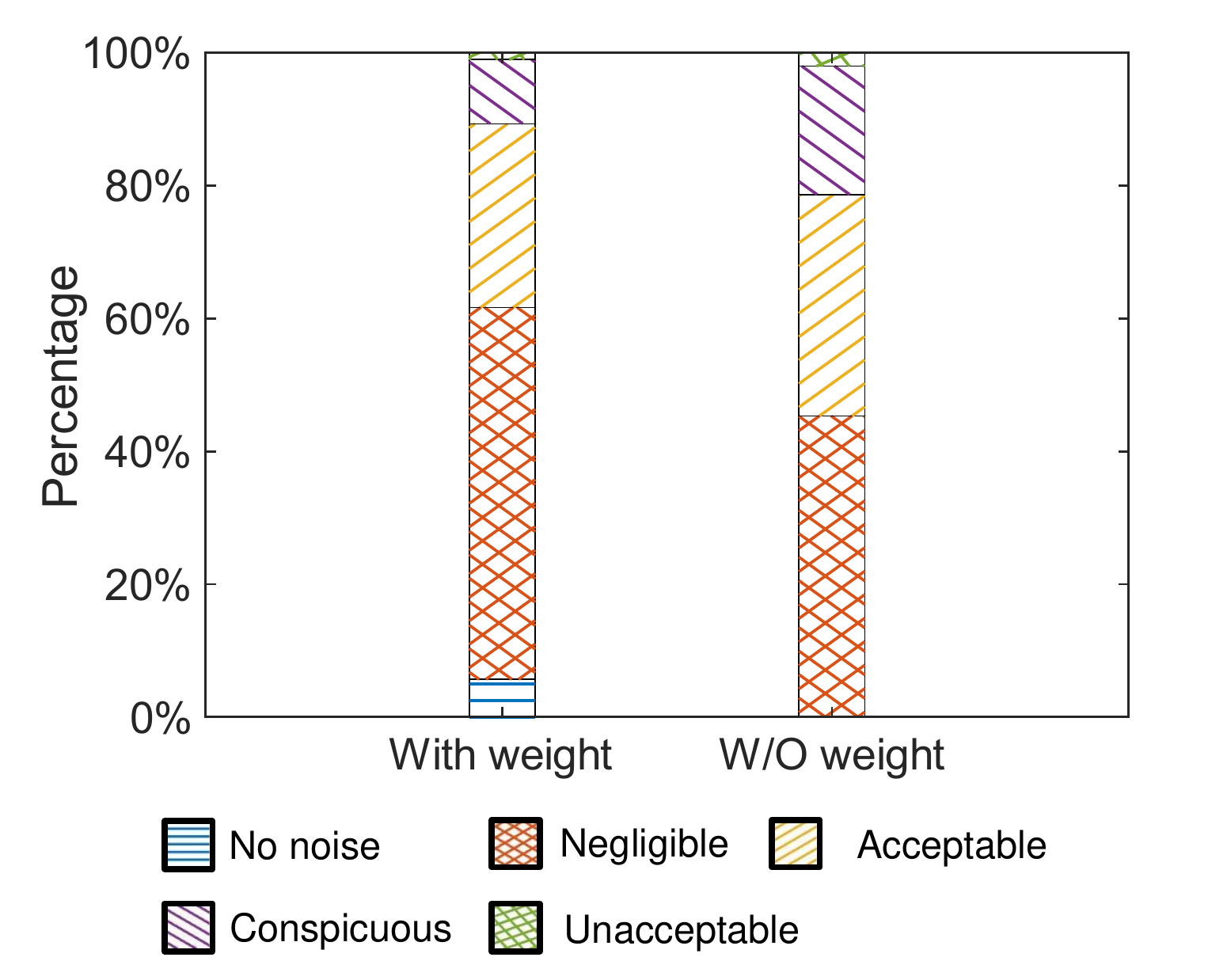}
\caption{Human perceptibility for \vm protected speech with and without weight-based constraint.}\label{fig: weight_hs}
\end{figure}
\end{document}

%% file: document/1_introduction.tex
\vspace{-5pt}
\section{Introduction}
\label{introduction}
The recent development in Artificial Intelligence (AI) has spurred the research on intelligent voice control. It is now possible to impersonate someone's voice almost impeccably, as shown in various ``deepfake" campaigns for misinformation and fraud~\cite{mahmud2021deep}. 
Deep learning based voice synthesis is the core technology used in deepfakes to generate speeches of arbitrary target speaker's voice without specific speech samples.
There are two types of speech synthesis approaches: text-to-speech (TTS) synthesis and voice conversion (VC).
TTS generates synthetic speeches using the script texts and a single speech sample from the speaker. 
However, 
it usually outputs unnatural and mechanical human speeches. 
VC, on the other hand,  directly 
transforms the voice of the original speaker to that of another speaker without altering the linguistic information within the speech.
VC is capable of generating vivid speech with real human speech characteristics, such as  pauses, moods, and even accents.

Advanced voice synthesis has been used in benign applications, such as reading stories to kids with their deceased grandma's voice~\cite{amazon2022grandma}, or letting your favorite movie star to guide your routes~\cite{voicenavigation}.
However, as shown in Fig.~\ref{fig: voice synthesis scenario}, human ears or automatic speaker verification (ASV) systems verify speaker identity through voice patterns, voice synthesis technology causes potential security threats.
First, voice assistants often require the authenticated voice to activate Siri or Google Assistant before users operate the locked smartphone.
Attackers could easily take control of the smartphone and send malicious commands with synthesized voice.
Second, ASV systems have been integrated in many popular Apps for user authentication.
For example, WeChat requires the user to speak a 6-digit number before the first-time login on an unseen device~\cite{wechatvoicelock}.
Such user verification is vulnerable to voice synthesis attacks.
Third, attackers can leverage synthetic voice to impersonate the victim in a fraud phone call~\cite{wang2022ghosttalk}.
It is difficult for human ears to differentiate high-quality synthetic speech from natural human voice~\cite{wenger2021hello}.

In response to the threat of voice synthesis attacks, researchers have developed synthetic human voice detection models to identify fake human speech~\cite{ahmed2020void, zhang2021one, li2021channel}. 
However, implementing voice synthesis detection on all mobile devices can be costly, and these detection methods may not achieve the claimed high accuracy in real-world scenarios~\cite{wenger2021hello}.
\rev{To defend against unauthorized voice synthesis attacks, Attack-VC~\cite{huang2021defending} adds adversarial perturbations to victim’s speech samples to disallow adversaries from generating synthetic voices from the victim.}
The synthesized speech derived from the protected speech will sound like a distinct voice from another predetermined target speaker rather than the victim speaker\footnote{``victim speaker" refers to the target of voice synthesis attack. ``target speaker" refers to a different speaker pre-determined by \vm to interfere with the voice synthesis.}.
\rev{Nevertheless, the application of Attack-VC is somewhat restricted. 
The PGD method used to optimize the adversarial perturbation requires the availability of the entire speech and a significant number of iterations to generate the adversarial speech samples. 
Therefore, the optimization process will not be able to keep up with the real-time speech. 
As a result, live human speech and voice signals in real-time applications, such as voice messages, online meetings, are still vulnerable to voice synthesis threats.}

In this work, we propose the \emph{first} real-time voice protection method, \vm, to protect the real-time speech in both digital and physical world.
Instead of optimizing the perturbation itself, we train a neural network to anticipate the most efficient perturbations in the future. 
As a result, the protective perturbation can be instantly incorporated into the real-time audio stream without causing extra time delay.
In addition, to protect the entire speech, we introduce universal perturbation headers to mask the beginning of the speech regardless of the speech content.
Based on human auditory perception characteristics, we introduce a supplementary weight-based amplitude constrained method to further alleviate the audio distortion of the protected speech.
In summary, the paper makes following contributions:
\vspace{-5pt}
\begin{itemize}
    \item We propose  \vm, a real-time protection system against voice synthesis attacks.
    By predicting the perturbation for upcoming streaming speech, \vm can shield the user's voice with negligible delays.
    \item We propose an optimization scheme to train universal perturbation to protect the exposed audio at the beginning of speech.
    This allows \vm to indiscriminately protect speech data with different lengths and sizes. We further introduce a weight-based amplitude constrained method to reduce human perceptibility for adversarial perturbations.
    \item We evaluate \vm performance on three different voice synthesis models. The experimental results show that \vm is capable of providing real-time defense against voice synthesis attacks on both digital and physical spaces, effectively shielding ASV systems and the human auditory system from malicious voice synthesis.
\end{itemize}

%% file: document/2_background.tex
\section{Background}
\label{background}
\subsection{Voice Synthesis}
\begin{figure}[t]
    \setlength{\abovecaptionskip}{0pt} 
    \setlength{\belowcaptionskip}{-18pt} 
    \centering
    \includegraphics[width=7.5cm]{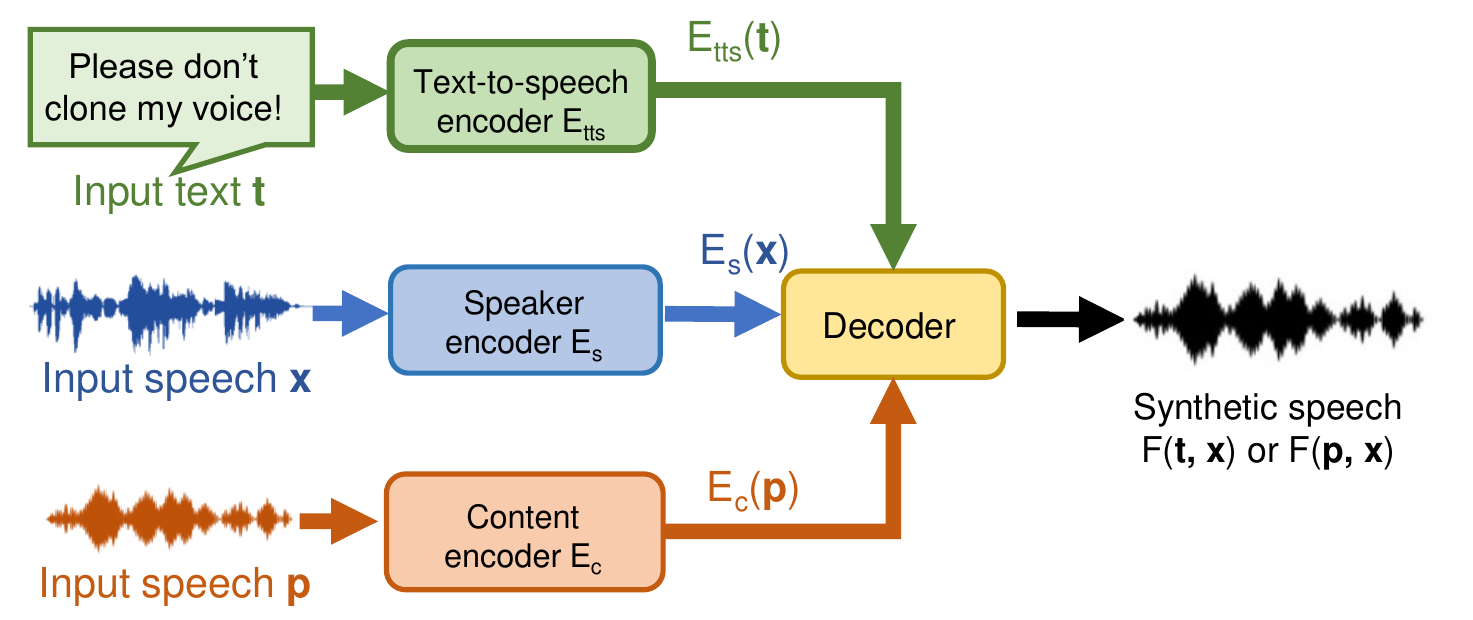}
    \caption{The speaker encoder in voice synthesis models is susceptible to adversarial attacks.}
    \label{fig:voice_synthesis}
\end{figure}

The purpose of voice synthesis attacks is to generate artificial speech resembling real human voices.
Fig.~\ref{fig:voice_synthesis} shows a general voice synthesis framework including both TTS and VC.
For TTS and VC models, the speech content embeddings are respectively extracted by the content encoders $E_{tts}$ or $E_{c}$ from text $\textbf{t}$ or speech $\textbf{p}$.
The speaker encoder $E_{s}$ encodes the victim speaker characteristics $E_{s}(\textbf{x})$, where $\textbf{x}$ is a speech sample from the victim speaker $x$.
Finally, the decoder will generate a synthetic speech $F(\textbf{x}, \textbf{p})$ or $F(\textbf{x}, \textbf{t})$ with voice features from an arbitrary victim speaker $x$.
\vspace{-4pt}

\subsection{Defense Against Voice Synthesis}
Attack-VC~\cite{huang2021defending} introduced \textit{embedding attack} that aims at the speaker encoder model $E_{s}$.
If we input an adversarial speech sample $\textbf{x} + \delta$ with little difference from $\textbf{x}$, the encoded vector $E_{s}(\textbf{x} + \delta)$ is close to another target speaker $y$ rather than the real victim speaker $x$. 
The adversarial attack for voice protection can be formulated in Problem~(\ref{embed_attack}) as follows:
\begin{equation}
\begin{split}
\mathop{\textnormal{minimize}}\limits_{\delta} \quad \mathcal{L}&(E_{s}(\textbf{x}+\delta), E_{s}(\textbf{y})) -\lambda \mathcal{L}(E_{s}(\textbf{x}+\delta), E_{s}(\textbf{x}))\\
\textnormal{subject to} \quad \Vert &\delta \Vert_{\infty} < \epsilon,
\end{split}
\label{embed_attack}
\end{equation}
where $\mathcal{L}$ is mean squared error (MSE) loss function, and $\textbf{y}$ is a speech sample from target speaker $y$.
The first term in Problem~(\ref{embed_attack}) minimizes the loss between the speaker embedding vector of adversarial input $E_{s}(\textbf{x}+\delta)$ and the target speaker $E_{s}(\textbf{y})$ to force the model to generate speech voice similar to the target speaker $y$, while the second term eliminates the victim speaker voice $E_{s}(\textbf{x})$.
Eventually, with protected speech input $x+\delta$, the synthetic voice cannot spoof the speaker verification or human ears anymore.
\rev{In Problem~(\ref{embed_attack}), it is necessary to set a predetermined target speaker $y$.
We run an additional experiment without setting a target speaker, (i.e., by setting $\lambda \rightarrow \infty$).
The result shows: despite that the synthesized speech becomes noisy, it still sounds like the voice from the victim speaker.}

\vspace{-5pt}
\subsection{Real-time Adversarial Attack}
Since human speech is a streaming audio signal, it is impractical to generate real-time adversarial perturbation using the PGD method. 
VoiceCamo~\cite{chiquier2021real} presents a real-time adversarial attack towards speech recognition (SR) models, which could forecast the upcoming adversarial perturbation by optimizing the model $g_{\theta}$.
The predictive model optimization in VoiceCamo is illustrated as follows:
\begin{equation}
\begin{split}
&\mathop{\textnormal{maximize}}\limits_{\theta} \ \mathbb{E}_{(x_{t}, y_{t})}[\mathcal{L}_{CTC}(\Bar{y}_{t}, y_{t})] \\ 
\textbf{s.t.} \quad \Bar{y}_{t} = &f_{\psi}(x_{t}+g_{\theta}(x_{t-r-\delta})) \quad \textnormal{and} \quad \Vert g_{\theta}(x_{t}) \Vert_{\infty} < \epsilon,
\end{split}
\label{rt_attack}
\end{equation}
where $\mathcal{L}_{CTC}$ is connectionist temporal classification (CTC) loss function, $x_{t}$ is a streaming speech audio, and $y_{t}$ is the true label of the speech transcript.
$f_{\psi}$ is the speech-to-text model with \emph{fixed} parameter $\psi$, $g_{\theta}$ is the predictive model to forecast the perturbation, and $x_{t-r-\delta}$ is the speech signal captured at time $t-r-\delta$.
After predicting the first perturbation, the input window slides to the next time slot to predict the second perturbation, and so on so forth.
Finally, the speech recognition accuracy will be significantly degraded when the real-time attack activates.

\begin{figure}[t]
    \setlength{\abovecaptionskip}{-2pt} 
    \setlength{\belowcaptionskip}{-20pt} 
    \centering
    \subfigure[Real-time communication applications.]{\includegraphics[width=0.35\textwidth]{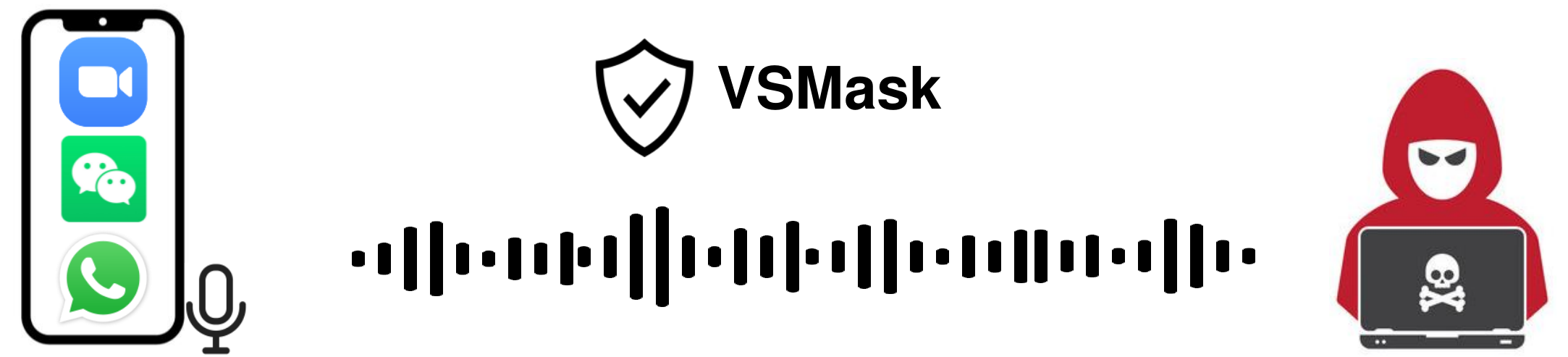}\label{fig:app1}}
    \subfigure[Social media platforms.]{\includegraphics[width=0.35\textwidth]{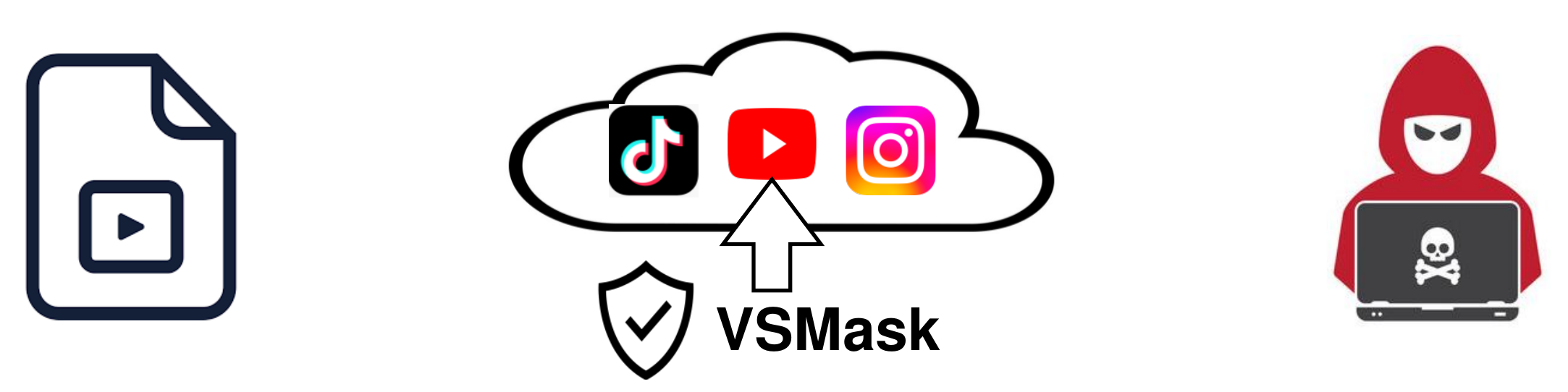}\label{fig: app2}}
    \subfigure[Real-world conversations.]{\includegraphics[width=0.35\textwidth]{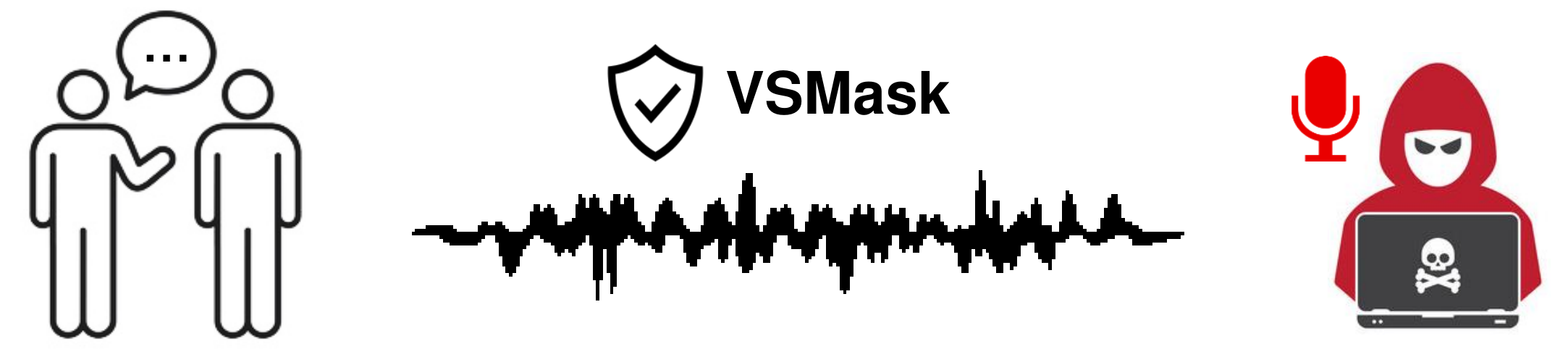}\label{fig: app3}}
    \caption{\vm provides protection in both online and offline scenarios.}
    \label{fig: vm scenario}
\end{figure}

\vspace{-4pt}
\section{Threat Model}
\label{threatmodel}
The state-of-the-art voice synthesis models allow the attackers to generate voice with a single speech sample from the victim speaker.
In this section, we define the threat model of voice synthesis attacks according to the adversary's knowledge about the victim and potential defenses.
\vspace{-6pt}
\subsection{Adversary Knowledge}
\vspace{-3pt}
The attacker can leverage voice synthesis models to generate artificial speech with the victim speaker's voice, in order to compromise the speaker verification systems or human ears.
The model could be well-trained or locally trained by the adversaries.
They can also test the synthesized speech on speaker recognition models to evaluate the voice similarity.
If they realize the samples are protected, they can apply denoising or other signal processing methods to recover the clear audio.
\subsection{Adversary Capability}
The adversary is able to save audio signals played in mobile devices, for example, smartphones or laptops, or record speech in physical world with a microphone.
The victim speaker's speech samples are available from multiple sources.

\noindent \textbf{Real-time Communication Applications.}
As shown in Fig.~\ref{fig:app1}, it is common for people to send voice memos or make voice calls using real-time communication apps. 
The attackers may hide in chat groups and record anyone's voice for voice synthesis.
Meanwhile, many people meet online since the COVID-19 pandemic.
The attackers can join public online meetings and record the victim's voice without seeking permission.
But existing defense methods fail to provide protection for the real-time speech. 

\noindent \textbf{Social Network Platforms.}
People are willing to share their daily life on social media.
There are billions of active users uploading their self-made videos to TikTok, YouTube and Instagram~\cite{usernumberins}.
Usually, all visitors have the access to download and reuse them.
Even with a short speech, malicious attackers could clone your voice and impersonate your identity.
Although existing defense methods are able to protect your voice on your social media, it requires long processing time before uploading~\cite{huang2021defending}.

\noindent\textbf{Real-world Conversation.}
In addition, real-world human speech is threatened by voice synthesis attacks.
While people are speaking in a presentation or conversation, an adversary can stealthily record the speech and launch voice synthesis attacks.
But existing defenses fail to mask our voice efficiently to allow the protection against malicious voice synthesis.
It is also challenging to physically inject perturbation into real speech.
\vspace{-8pt}
\subsection{Defender Knowledge}
\rev{Meanwhile, we consider a white-box scenario where the defenders have complete knowledge of the adversarial voice synthesis model, including model structure and model parameters. }
To optimize \vm, defenders also need to collect a few clear speech samples from the protected victim speaker $x$.
In addition, defenders have multiple target speaker candidates $y_{i}$ speech samples from public datasets, and they can select the best target $y$ for the victim speaker $x$ to maximize the voice disparity. 
Raw speech audios will be locally processed by \vm and uploaded to real-time applications or social media platforms without extra time delay.
\vspace{-2pt}

%% file: document/3_model.tex
\vspace{-8pt}
\section{\vm System Design}
\label{system_design}
\vspace{-3pt}
\subsection{Design Consideration}
One typical attack approach towards
ASR system is to hide the adversarial perturbations in the psycho-acoustic features of human speech~\cite{schonherr2018adversarial}, which could be a potential solution for voice synthesis defense.
However, these perturbations require the availability of the entire speech to mask them. As a result, it cannot be applied in real-time applications.

We also notice that universal perturbations can fool the ASR system regardless of the original speech contents~\cite{lu2021exploring}.
But such short perturbations are too short to protect the long speech from voice synthesis attacks.
To extend the protection scope, we attempt to leverage periodical universal perturbation to shield human speech.
Unfortunately, the performance of periodical perturbation method is unsatisfying.
The defense is effective only when the perturbation signal $\delta$ is strong.
Generally, short universal perturbations have limited degrees of freedom, which limits the protection performance.z
In contrast, longer perturbation period usually causes over-fitting;
besides, periodical perturbation can be denoised by attackers once they capture the perturbation at the interval of the speech.
Thus, it is crucial to develop a robust and real-time protection mechanism to safeguard against voice synthesis attacks.
\vspace{-6pt}
\subsection{Real-time Predictive Model}
\label{predictivemodel}
\begin{figure}[t]
    \setlength{\abovecaptionskip}{0pt} 
    \setlength{\belowcaptionskip}{-15pt} 
    \centering
    \includegraphics[width=7.5cm]{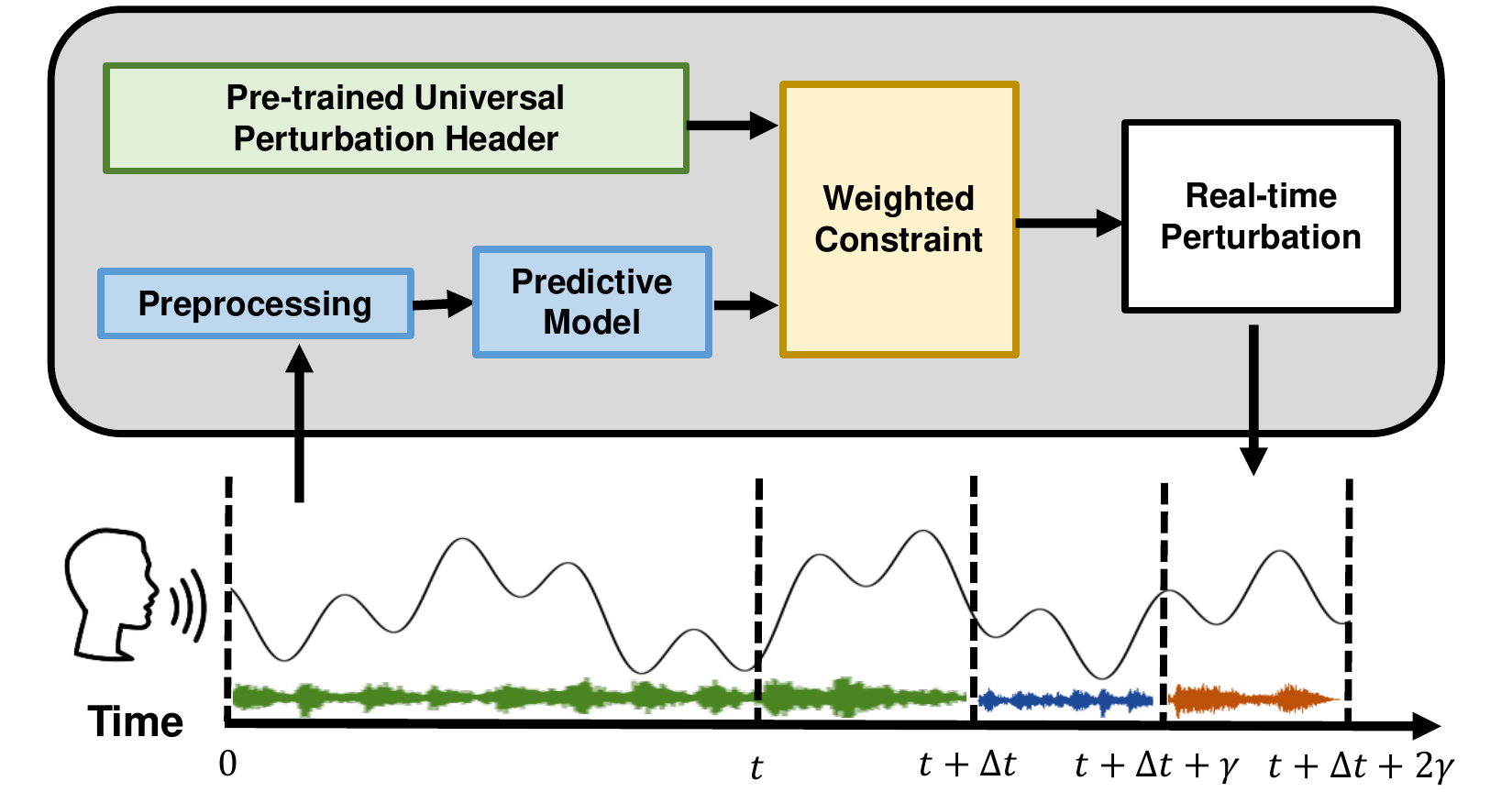}
    \caption{\vm system framework.}
    \label{fig:vm_system}
\end{figure}
\noindent \textbf{Preprocessing.}
Fig.~\ref{fig:vm_system} shows the system framework of \vm.
The input of \vm is a streaming human speech signal, which is vulnerable to voice synthesis attacks.
Once \vm detects human speech starts, it will capture the upcoming audio signal.
\rev{During preprocessing, the audio signal is transformed into a mel-spectrogram, with the frequencies converted to the non-linear mel scale to match human hearing. } 
The resulting mel-spectrogram has the same dimensions as the input of the speaker encoder $E_{s}$ in the targeted voice synthesis model.

\noindent \textbf{Model Architecture.}
Next, the converted mel-spectrogram works as the input of \vm predictive model $f_{\theta}$.
The output of the predictive model is also a mel-spectrogram with shorter length.
\rev{Learned from VoiceCamo~\cite{chiquier2021real}, \vm selects a real-time predictive model  composed of 7 down-sampling blocks and 5 up-sampling blocks, which outperforms models with less layers and maintains similar performance as models with more layers.}
Each down-sampling block contains a reflection padding followed by a 2-D convolution layer, a 2-D batch form, and a PReLU activation function;
each up-sampling block consists of a 2-D ConvTranspose layer and a leaky ReLU activation function.
The final up-sampling block is followed by a tanh function to constrain $\Vert \delta \Vert_{\infty}$. 
The output mel-spectrogram will be converted to time-domain audio signal and injected into the raw audio.
\newrev{We illustrate the fine-tuning process and model architecture details in Appendix~\ref{model_appendix}. }

\noindent \textbf{Optimization and Workflow.}
Suppose the speaker encoder function in the target voice synthesis model is $E_{s}$, we could formulate predictive model training as the following optimization problem:
\begin{equation}
\begin{split}
\mathop{\textnormal{minimize}}\limits_{\theta}& \quad \mathcal{L}(E_{s}(\textbf{x}+\delta), E_{s}(\textbf{y})) -\lambda \mathcal{L}(E_{s}(\textbf{x}+\delta), E_{s}(\textbf{x})),\\
\textnormal{subject to}& \quad  \delta^{\gamma}_{t+\Delta t} = f_{\theta}(\textbf{x}_{t})\quad \textnormal{and} \quad \Vert \delta \Vert_{\infty} < \epsilon,
\end{split}
\label{real-time optimization}
\end{equation}
where $\mathcal{L}$ is MSE loss function, $\textbf{x}$ is the input streaming speech from the victim speaker $x$, $\textbf{y}$ is a speech sample from the selected target speaker $y$ to mislead $E_{s}$, and $\delta$ is the perturbation predicted by model $f_{\theta}$.
$\delta^{\gamma}_{t+\Delta t}$ is the perturbation added at $t+ \Delta t$ with length $\gamma$.
\rev{Since the predictive model $f$ requires an input signal $\textbf{x}_{t}$ with a short length t,  \vm does not require the entire speech in advance, facilitating real-time protection.}

As illustrated in Figure~\ref{fig:vm_system}, \vm initially generates a perturbation of length $\gamma$ (represented by the blue waveform) after a short time delay of $\Delta t$. Subsequently, the input window is slid to the next time slot from  $\gamma$ to $t+\gamma$ to generate the second perturbation (represented by the red waveform), and this process repeats. 
\rev{In contrast to Attack-VC, \vm fine-tunes the predictive model parameters $\theta$ rather than manipulating the perturbation $\delta$. As a result, once the predictive model $f_{\theta}$ is established, \vm can identify the best perturbation $\delta$ in a single computation step, rather than undergoing a large number of iterations.}

\vspace{-5pt}
\subsection{Universal Perturbation Header}
\vspace{-3pt}
However, \vm predictive model has some deficiencies, since the voice synthesis models can generate synthetic speech via very short speech samples from the victim speaker.
For example, the victim speaker gives a speech example -- \emph{``please protect my speech from voice synthesis"}.
The predictive model could well protect the trailing phrases \emph{``my speech from voice synthesis"}.
Yet, the leading words of the speech \emph{``please protect"} are still exposed to the attackers, as 
\vm prediction model cannot generate perturbation for the beginning of the speech.
Therefore, the attacker may be able to clip the speech audio and synthesize speech with an audio segment.

To address this challenge, we resort to a universal perturbation added at the beginning of the streaming speech signal to protect the entire speech.
Recent work~\cite{lu2021exploring} shows that regardless of the utterance, there exists universal perturbations that can force the ASR system to output specific transcripts, e.g., ``yes" or ``no". 
This universal perturbation maintains high success rate especially when the input speech is short.
Fortunately, the ``blind spot" of the predictive model is also small.
Therefore, it is feasible to train a universal perturbation forcing the speaker encoder to output the selected target speaker $y$ embedded vector $E_{s}(\textbf{y})$ rather than the real vector $E_{s}(\textbf{x})$, regardless of the speech contents.

In order to train such a universal perturbation header $\delta_{h}$, we clip the training speech samples from the victim speaker $x$ to audio clips $\textbf{x}_{i}$ with the same length $t+\Delta t$.
Then, we use PGD method to optimize the universal perturbation header $\delta_{h}$ as follows:
\begin{equation}
\begin{split}
&\mathop{\textnormal{minimize}}\limits_{\delta_{h}} \quad \sum_{\textbf{x}_{i} \in \mathcal{D}} \mathcal{L}_{s}(\delta_{h}, \textbf{x}_{i}, \textbf{y}),\\
\textnormal{subject to} &\quad \mathcal{L}_{s}(\delta_{h}, \textbf{x}_{i}, \textbf{y}) = \mathcal{L}(E_{s}(\textbf{x}_{i}+\delta_{h}), E_{s}(\textbf{y})) \\
&-\lambda \mathcal{L}(E_{s}(\textbf{x}_{i}+\delta_{h}), E_{s}(\textbf{x}_{i})) \quad \textnormal{and} \quad \Vert \delta_{h} \Vert_{\infty} < \epsilon.\\
\end{split}
\label{perturbation header}
\end{equation}
where $\mathcal{D}$ is the collection of clipped speech $\textbf{x}_{i}$.

\begin{figure}[t]
    \setlength{\abovecaptionskip}{-0pt} 
    \setlength{\belowcaptionskip}{-5pt} 
    \centering
    \subfigure[The spectrogram of the natural speech $\textbf{x}$.]{\includegraphics[width=0.225\textwidth, height = 0.05\textwidth]{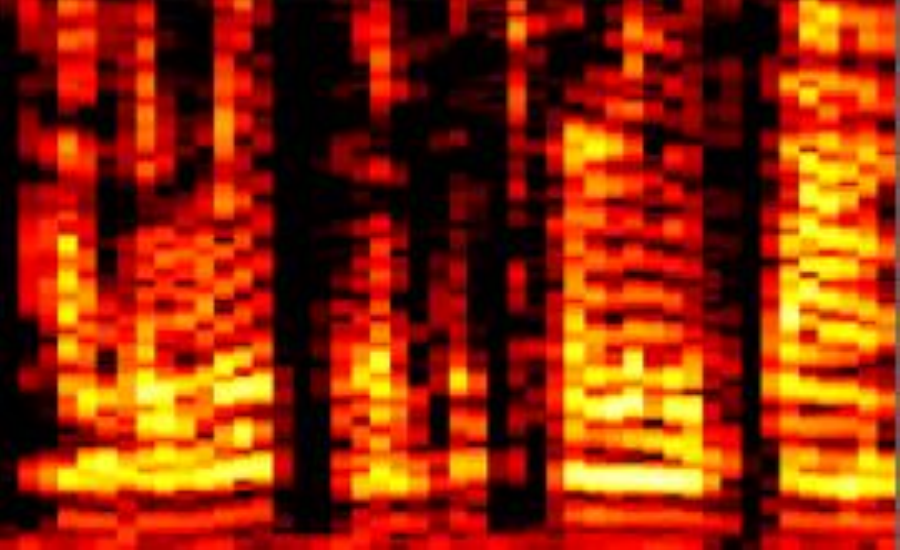}\label{fig:sample1}}\quad
    \subfigure[The spectrogram of the protected speech $\textbf{x}+\delta$.]{\includegraphics[width=0.225\textwidth, height = 0.05\textwidth]{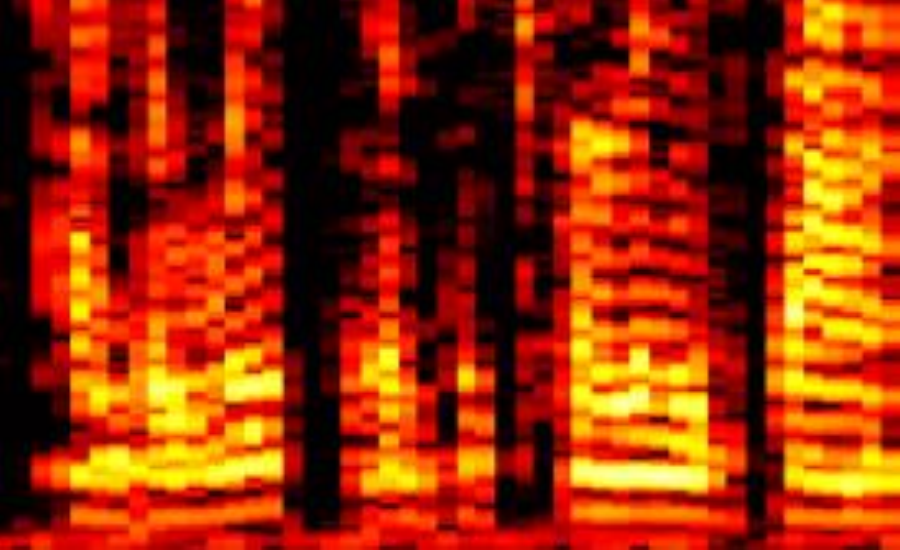}\label{fig:sample2}}
    \subfigure[The spectrogram of synthetic speech $F(\textbf{x})$.]{\includegraphics[width=0.225\textwidth, height = 0.05\textwidth]{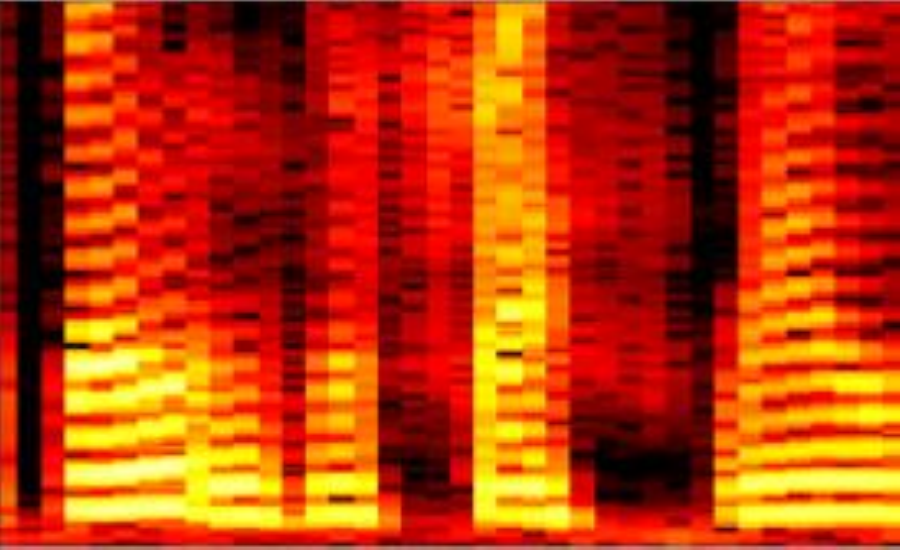}\label{fig:sample3}}\quad
    \subfigure[The spectrogram of synthetic speech $F(\textbf{x}+\delta)$.]{\includegraphics[width=0.225\textwidth, height = 0.05\textwidth]{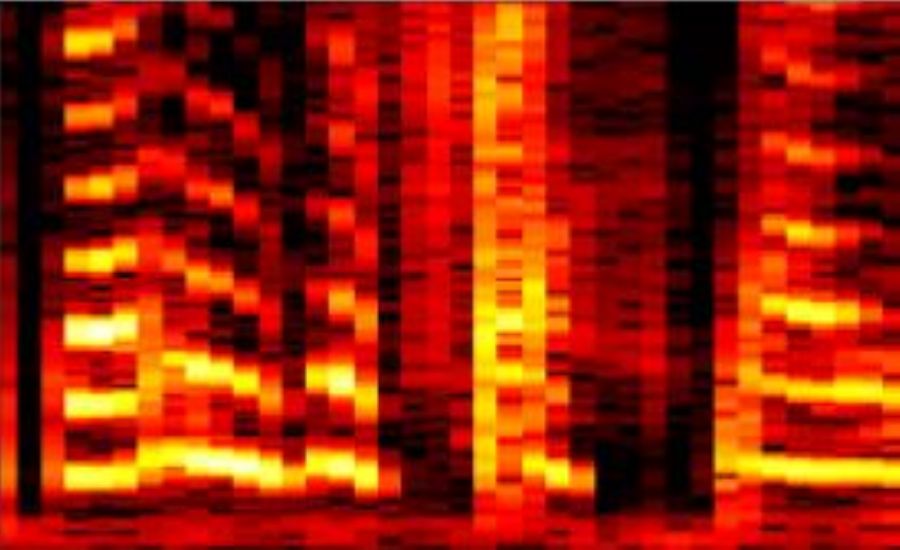}\label{fig:sample4}}
    \caption{Speech protected by \vm effectively degrades the voice synthesis model performance.}
    \vspace{-10pt}
    \label{fig: examples}
\end{figure}

Before the optimization process in Section~\ref{predictivemodel}, we add the universal header into all training samples to enhance the protection performance.
After combining the predictive model and universal header, \vm successfully protects the streaming speech in real-time scenarios.
Fig.~\ref{fig:sample1} shows a natural speech sample $\textbf{x}$ exposed to the attackers.
Fig.~\ref{fig:sample2} is the audio spectrogram of protected speech $\textbf{x}+\delta$.
The perturbation causes little difference between these two speech samples.
Fig.~\ref{fig:sample3} and Fig.~\ref{fig:sample4} are respectively the synthetic speech spectrograms from $\textbf{x}$ and $\textbf{x}+\delta$ with the same speech content.
Even though $\textbf{x}$ and $\textbf{x} + \delta$ are very similar, the synthetic speech $F(\textbf{x})$ shows male voice patterns whereas $F(\textbf{x}+\delta)$ presents female voice with higher frequencies.

\vspace{-10pt}
\subsection{Weight-based Noise Mitigation}
\vspace{-3pt}
Inevitably, the injected protective perturbation will cause audio distortion.
Prior work usually constrains the perturbation amplitude to minimize the human perception, or considers the amplitude in the optimization to achieve the best trade off between noise level and performance.
However, even as the perturbation is well-optimized, there is still significant background noise in the speech signal.

\begin{figure}[t]
    \setlength{\abovecaptionskip}{0pt} 
    \centering
    \includegraphics[width=0.33\textwidth]{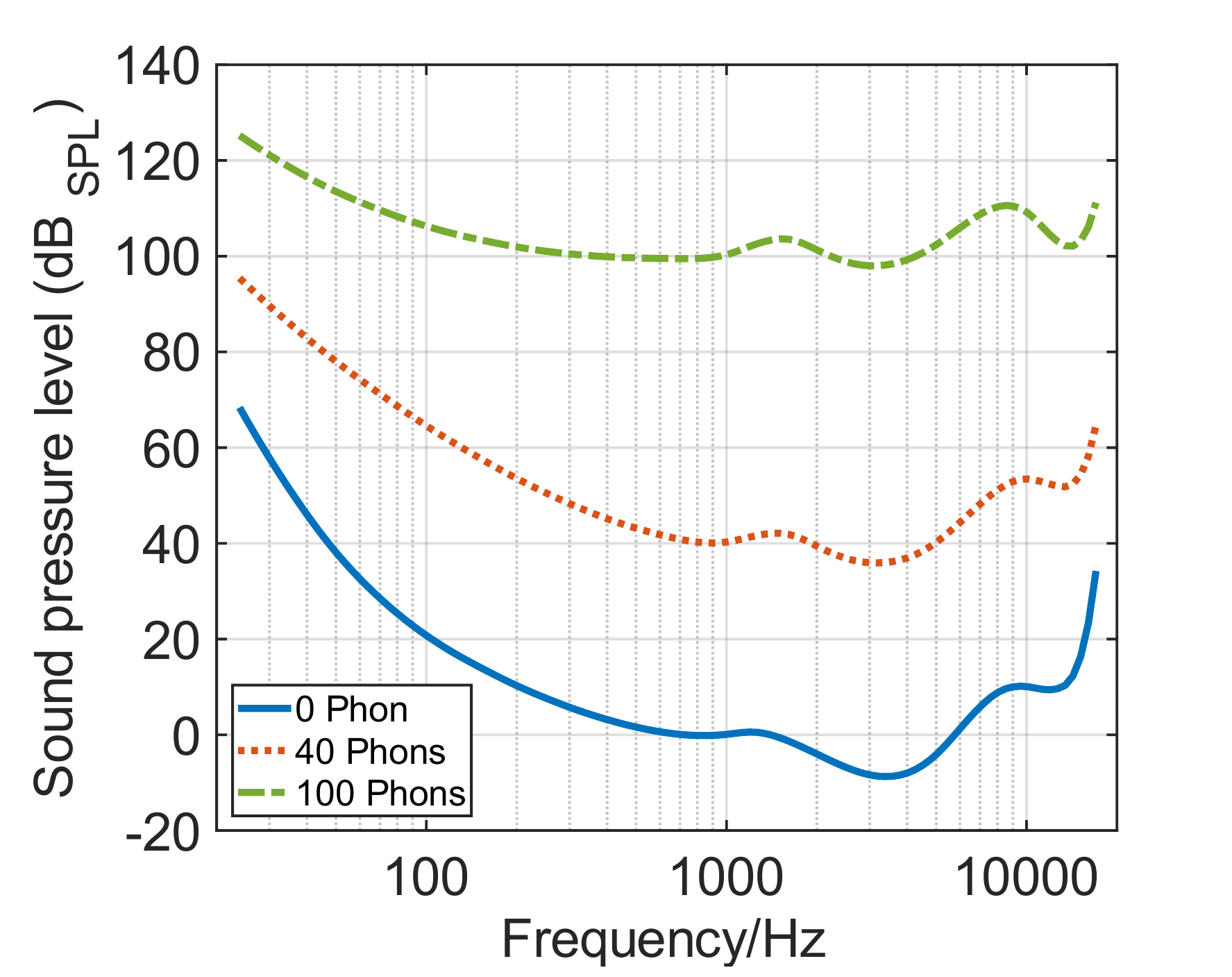}
    \caption{Equal-loudness contour.}
    \label{fig:equalloudness}
\end{figure}

To degrade the perturbation perceptibility of \vm, we look into human hearing sensitivity.
Fig.~\ref{fig:equalloudness} shows equal-loudness contour~\cite{equalcontour}, a measurement of human hearing perception under different frequency tones.
For low frequency ($< 1.6$ kHz) or high frequency ($> 4$ kHz), higher sound pressure is required to cause the same loudness level in human ears.
For medium frequency band between $1.6$ kHz and $4$ kHz, human ears have the highest sensitivity.
Based on this phenomenon, one potential approach is to hide all perturbations in the low frequency band.
However, our experiment shows that it fails to provide sufficient protection for our speeches.

Therefore, we need to inject a global perturbation into the entire spectrogram.
To mitigate human hearing perception, we are supposed to reduce the perturbation amplitude between $1.6$ kHz and $4$ kHz.
However, it is expected that reducing the perturbation amplitude also degrades the protection performance.
To compensate the scarcity in medium frequency, we increase the amplitudes in other frequency bands which are hard to sense for human ears.
Therefore, we rewrite the amplitude constraint as follows:
\vspace{-5pt}
\begin{equation}
\begin{split}
\delta &= \begin{bmatrix} \delta_{l} &  \delta_{m} & \delta_{h} \end{bmatrix}^{T},\\
\textbf{s.t.} \quad \Vert \delta_{l} \Vert_{\infty} &< \epsilon_{1}, \ \Vert \delta_{m} \Vert_{\infty} < \epsilon_{2}, \ \Vert \delta_{h} \Vert_{\infty} < \epsilon_{3},
\end{split}
\label{weight}
\end{equation}
where $\epsilon_{2}<\epsilon_{3}<\epsilon_{1}$.
By adding such weight-based constraint on the perturbation amplitude, we successfully improve the protected audio quality by limiting the human perception of perturbation without sacrificing the protection performance.
\vspace{-10pt}
\subsection{\vm Application Scenarios}
As we mentioned in Section~\ref{threatmodel}, the attackers have many different approaches to obtain the speech audio from the victim speaker.
Here, we describe the versatile application scenarios of \vm.

\noindent \textbf{Online Protection.}
The first application is to protect human voice in real-time communication and social media Apps. 
For these Apps, \vm can work as a plug-in component to
protect the user's voice.
Once \vm detects the start of the speech, it will automatically add the universal perturbation header, and concurrently capture the speech audio.
After injecting the perturbation header, \vm has predicted the following perturbation for the next time slot, and then it iteratively protects the upcoming speech stream.
Based on the prediction mechanism, \vm causes no extra time delay on the real-time communication.

\noindent \textbf{Real-world Physical Protection.}
In real-world scenarios, the attackers may stealthily record the victim's voice and implement voice synthesis attack.
To counter such adversaries, users can protect themselves by installing \vm on their mobile devices.
While talking, \vm will play perturbation noise signal and effectively inject adversarial speech samples in the eavesdropped audio.
In the end, it will protect our real-world speech against voice synthesis attack, as shown in the experiments in Section~\ref{discussion}. 


%% file: document/4_evaluation.tex
\vspace{-10pt}
\section{Evaluation}\label{sec:evaluation}
\subsection{Target Models and Experiment Design}
\begin{table*}
\caption{We compare \vm defense performance with other online or offline baseline defense methods. \vm outperforms all other real-time protection methods and achieves similar performance as offline PGD method. }
\vspace{0pt}
\label{Tab:overall performance}
\normalsize
\centering
\begin{center}
\begin{tabular}{cccccccccc}
\toprule
Type                           & Method                  & \multicolumn{3}{c}{Score}                        & \multicolumn{2}{c}{Accuracy}   & \multicolumn{3}{c}{STOI}                          \\ 
\midrule
\multirow{2}{*}{Non-synthetic} & Raw speech              & \multicolumn{3}{c}{$0.780$}                      & \multicolumn{2}{c}{$98.8\%$}    & \multicolumn{3}{c}{1.000}                           \\ 
\cmidrule{2-10}
                              & Protected speech        & \multicolumn{3}{c}{$0.429$}                      & \multicolumn{2}{c}{$68.5\%$}    & \multicolumn{3}{c}{0.839}                         \\ 
\bottomrule
\toprule
\multirow{2}{*}{Type}          & \multirow{2}{*}{Method} & \multicolumn{2}{c}{M2M}         & \multicolumn{2}{c}{F2F}         & \multicolumn{2}{c}{M2F}         & \multicolumn{2}{c}{F2M}          \\ 
\cmidrule{3-10}
                               &                         & Score          & ASR      & Score          & ASR      & Score          & ASR      & Score          & ASR       \\ 
\midrule
\multirow{5}{*}{Synthetic}     & Raw speech              & $0.595 $       & $91.9\%$       & $0.612 $       & $93.2\%$       & $0.561$        & $88.3\%$       & $0.546 $       & $86.0\%$        \\ 
\cmidrule{2-10}
                               & Random noise           & $0.516 $       & $86.6\%$       & $0.538 $       & $89.0\%$       & $0.505 $       & $84.0\% $      & $0.473$        & $81.5\%$        \\ 
\cmidrule{2-10}
                               & Periodical perturbation & $0.192 $       & $11.0\%$       & $0.203 $       & $12.5\%$       & $0.177 $       & $9.8\%$       & $0.156 $       & $8.6\%$        \\ 
\cmidrule{2-10}
                               & Offline PGD             & 0.064          & $0.0\%$        & 0.085          & $0.0\%$        & 0.049         & $0.0\%$        & 0.055         & $0.0\%$         \\ 
\cmidrule{2-10}
                               & \textbf{\vm}               & \textbf{0.077} & \textbf{0.0\%} & \textbf{0.104} & \textbf{0.0\%} & \textbf{0.056} & \textbf{0.0\%} & \textbf{0.073} & \textbf{0.0\%}  \\
\bottomrule
\end{tabular}
\end{center}
\vspace{2pt}
\begin{tablenotes}\footnotesize
\centering
\item M2M: Male-to-Male, F2F: Female-to-Female, M2F: Male-to-Female, F2M: Female-to-Male.
\end{tablenotes}
\vspace{-8pt}
\end{table*}

\noindent\textbf{Target Voice Synthesis Model.}
Moreover, we select the following three representative voice synthesis models to evaluate \vm performance:
\vspace{-5pt}
\begin{itemize}
    \item AdaIN-VC~\cite{chou2019one}: an unsupervised one-shot voice conversion model based on instance normalization; 
    \item AutoVC~\cite{qian2019autovc}: a new style auto encoder with carefully designed bottleneck, which is the first work achieving zero-shot voice conversion with non-parallel speech data;
    \item SV2TTS~\cite{jia2018transfer}: a TTS model that converts text transcripts to human speech with the target speaker voice. 
\end{itemize}
\vspace{-2pt}
\noindent \textbf{Dataset.}
For AdaIN-VC and AutoVC models, we use CSTR VCTK Corpus~\cite{veaux2017cstr} dataset, a widely used speech dataset in the voice conversion studies, to evaluate the performance of \vm.
VCTK includes speech samples from 109 English speakers with different accents.
Each speaker reads up to 400 sentences randomly selected from newspapers or elicitation.
For SV2TTS model, we select LibriSpeech dataset~\cite{panayotov2015librispeech} for evaluation.
In particular, as some speech samples are too short to activate \vm predictive model, we only use speech audios longer than 3 seconds for training and testing.

\vspace{-6pt}
\subsection{\vm Effectiveness on ASV System}

\label{sec:vm_asv}
\noindent\textbf{ASV System for Evaluation.}
To evaluate \vm effectiveness of reducing the similarity between the synthetic speech voice and victim speaker's voice, we utilize SpeechBrain~\cite{speechbrain}, a state-of-the-art model to calculate the voice similarity of two speech samples. 
SpeechBrain speaker verification model is composed of emphasized channel attention, propagation and aggregation in time delay neural network (ECAPA-TDNN) to extract speaker embeddings.
The similarity scores are calculated by cosine similarity of embedding vectors of two speech samples.
The higher the score is, the greater the possibility that two speech samples are from the same speaker.
If the score is higher than a threshold $k$, the speech can pass the verification.
In our experiment, we set the default threshold $k=0.25$, which achieves the best accuracy for clear human speech samples.

\noindent \textbf{\vm and Baseline Methods Setup.}
In this section, we introduce \vm evaluation setup along with 4 other baseline methods for comparison.
\rev{We evaluate the methods in the digital domain by injecting perturbations into digital audio signals.}

\noindent \emph{\vm}: 
\rev{First, for \vm method, we set $\lambda = 1$ in Problem (\ref{real-time optimization}) and (\ref{perturbation header}) according to the $\lambda$ value in~\cite{wenger2021hello}, the input time window length $t$ as $1.25$ seconds.}
To ensure real-time prediction, it is important to avoid setting a very short time-delay. 
On the other hand, a long time-delay can lead to a degradation in the performance of \vm. Therefore, we have set both the time-delay ($\Delta t$) and output length ($\gamma$) to be 0.4 seconds.
Therefore, the perturbation header length is $t+\Delta t = 1.65$ seconds.
Moreover, we set the perturbation amplitude constraint $\epsilon=0.10$.
The perturbation amplitude after noise mitigation approach is: 
$\epsilon_{3}=\epsilon$, $\epsilon_{1}=1.15~ \epsilon$ and $\epsilon_{2}=0.85~ \epsilon$.

\rev{In addition, we randomly select 30 male and 30 female speakers from the dataset.
We use AdaIN-VC model to generate 100 synthesized speech samples for each speaker.}
For female victim speakers, we select a male speaker as the target speaker, and vice versa.
For voice conversion models, the gender of the speaker matters, since male and female have different voice patterns, e.g., female speakers usually have higher frequency utterances. 
We split the results according to different genders of victim speaker and speech source speaker.
\vspace{2pt}

\noindent\emph{Random Noise:}
One possible way to degrade the voice synthesis model performance is to add random noise into the speech signal $\textbf{x}$.
In the experiment, we set the random noise amplitude $\epsilon_{n}=0.15$.\\
\emph{Periodical Perturbation:}
Since the universal header can protect short speech regardless of the speech content, periodical universal perturbation is a potential solution for real-time defense.
In the experiment, we recurrently inject the same universal perturbation header with amplitude $\epsilon_{p}=0.12$ into the raw speech audio and evaluate its protection performance.
\vspace{2pt}


\noindent\emph{Offline PGD Method:}
In addition, we compare our method with offline PGD method in Attack-VC~\cite{huang2021defending}.
We apply the embedding attack method as shown in Problem (\ref{embed_attack}), while the iteration number is $1,500$, $\epsilon_{off} = 0.10$.
Even though offline PGD method is not feasible for real-time protection, it can help to find the upper bound of protection performance.
\vspace{2pt}

\noindent \textbf{\vm Evaluation on ASV System.}
We quantitatively evaluate the ASV system performance with non-synthetic speech samples.
For the clear audios without \vm protection, the ASV system achieves accuracy above 98\%.
But for speech audios protected by \vm, similar to the result in Attack-VC, the ASV accuracy degrades since the perturbation distorts the voice patterns in the speech.
In this way, \vm may affect speaker recognition functions in real-world scenario such as voice assistant activation.
\rev{Fortunately, mobile devices can cancel out the perturbation from loudspeakers using echo cancellation~\cite{murano1990echo}. 
So \vm will not interfere with the activation process.}
In addition, we use Short-Time Objective Intelligibility (STOI)~\cite{taal2011algorithm} to measure the speech quality of protected audio.
The result shows that the speech under \vm protection is fully intelligible for human ears.

Next, we compare \vm with different baseline defense methods, where the target voice synthesis model is AdaIN-VC.
We use attack success rate (ASR) to evaluate the performance of voice synthesis attacks.
For reference, when the attacker clones voice with raw speech audios, more than 85\% synthetic audios can pass the ASV system with high similarity score.
While the voice conversion is intra-gender (male-to-male or female-to-female), the ASR is even higher.
If the victim speaker is female, voice conversion shows slightly higher ASR than male victim speaker.
This result confirms the effectiveness of compromising ASV systems with voice synthesis attacks. 

Then, we evaluate the baseline real-time defense methods.
We first synthesize voice from speech with white noise.
When the perturbation amplitude is small ($\epsilon_{n}=0.15$), random noise cannot deteriorate the voice synthesis performance.
Actually, the embedding speaker vectors $E_{s}(\textbf{x})$ and $E_{s}(\textbf{x}+\delta)$ show little difference when $\delta$ is weak random noise.
If stronger noise is implemented, it works because of poor speech quality.
When the noise amplitude $\epsilon > 0.4$, no synthetic speech samples can compromise ASV model.
But the protected audio $\textbf{x}+\delta$ is illegible for human ears.
Consequently, strong noise is not a feasible protection method in real world.

\vspace{2pt}
Second, we use the periodical universal perturbation for defense.
As shown in Table~\ref{Tab:overall performance}, periodical universal perturbation significantly reduces the similarity of the synthetic voice and the victim's voice.
But there are still more than 10\% intra-gender synthetic audios that could pass ASV systems.
The reason is that when perturbation amplitude is small, universal perturbation cannot compromise the entire speech, especially for long speech samples.
We can increase the amplitude or cycle length to improve the periodical method performance.
However, if the perturbation is too strong, it can lead to significant audio distortion, and using a long period for perturbation may not be effective in protecting short speech segments.
\begin{figure}[tbp]
    \setlength{\abovecaptionskip}{0pt} 
    \centering
    \subfigure[The similarity scores for different voice synthesis models]{\includegraphics[width=4.1cm]{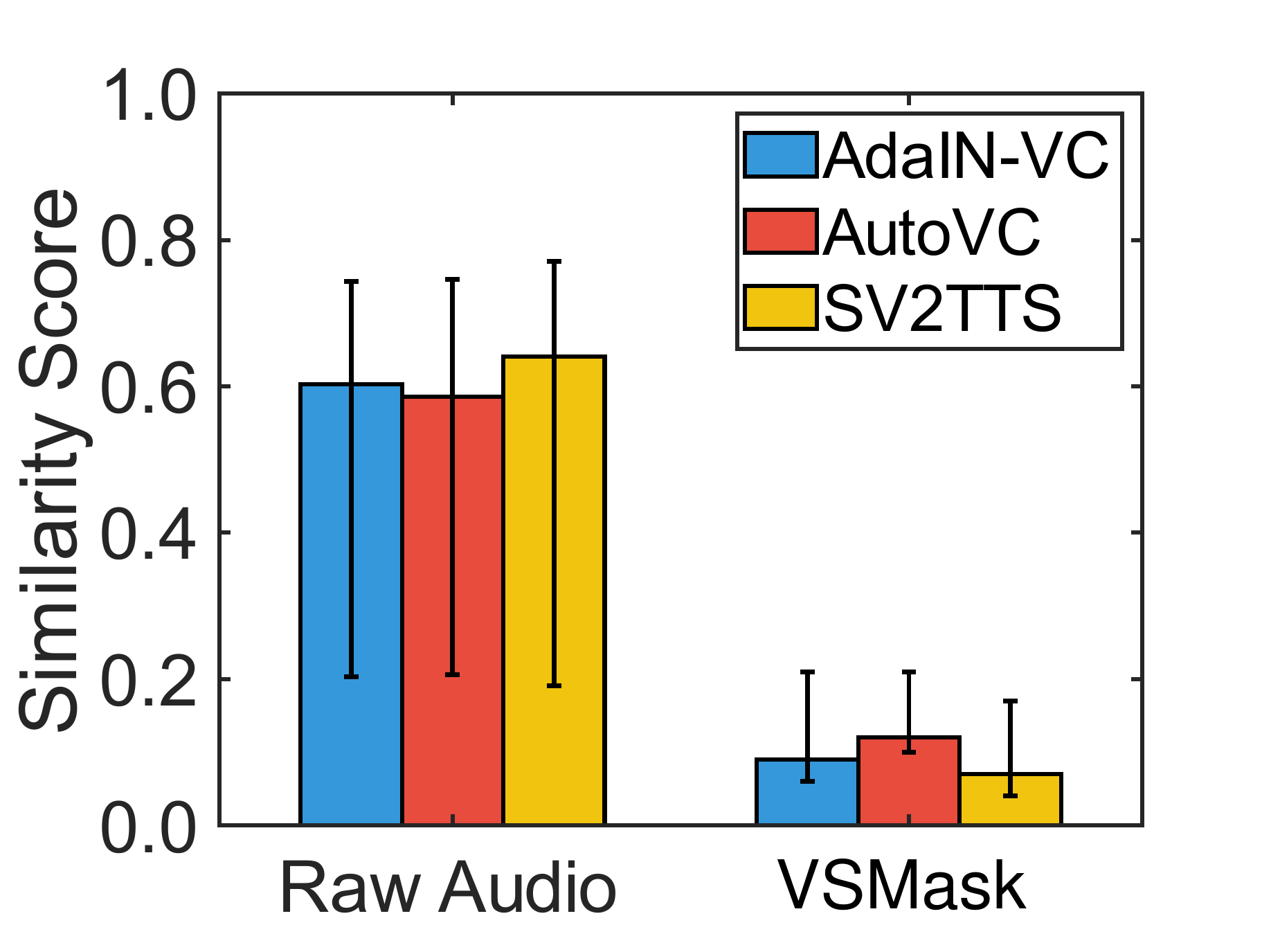}\label{fig:score}}
    \
    \subfigure[The ASR for different voice synthesis models]{\includegraphics[width=4.1cm]{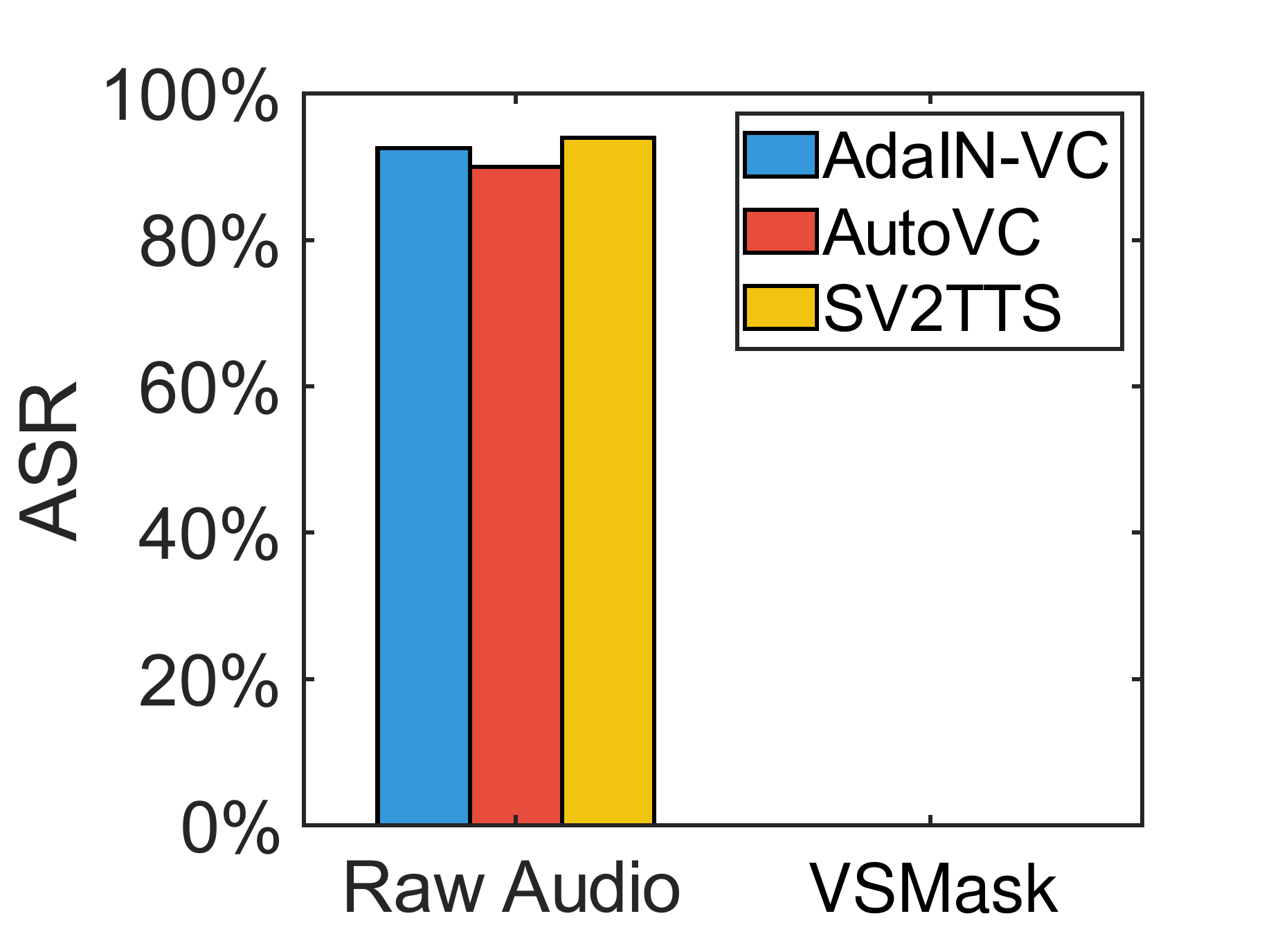}\label{fig: rates}}
    \caption{For all target voice synthesis models, \vm significantly reduce the voice similarity. With \vm protection, no synthetic speech can pass the speaker verification system.}
    \label{fig: 3_models_performance}
\end{figure}

\begin{figure*}[t]
    \setlength{\abovecaptionskip}{0pt} 
    \setlength{\belowcaptionskip}{0pt}
    \centering
    \subfigure[The noise level of protected speech for voice conversion models. ]{\includegraphics[height=4.5cm]{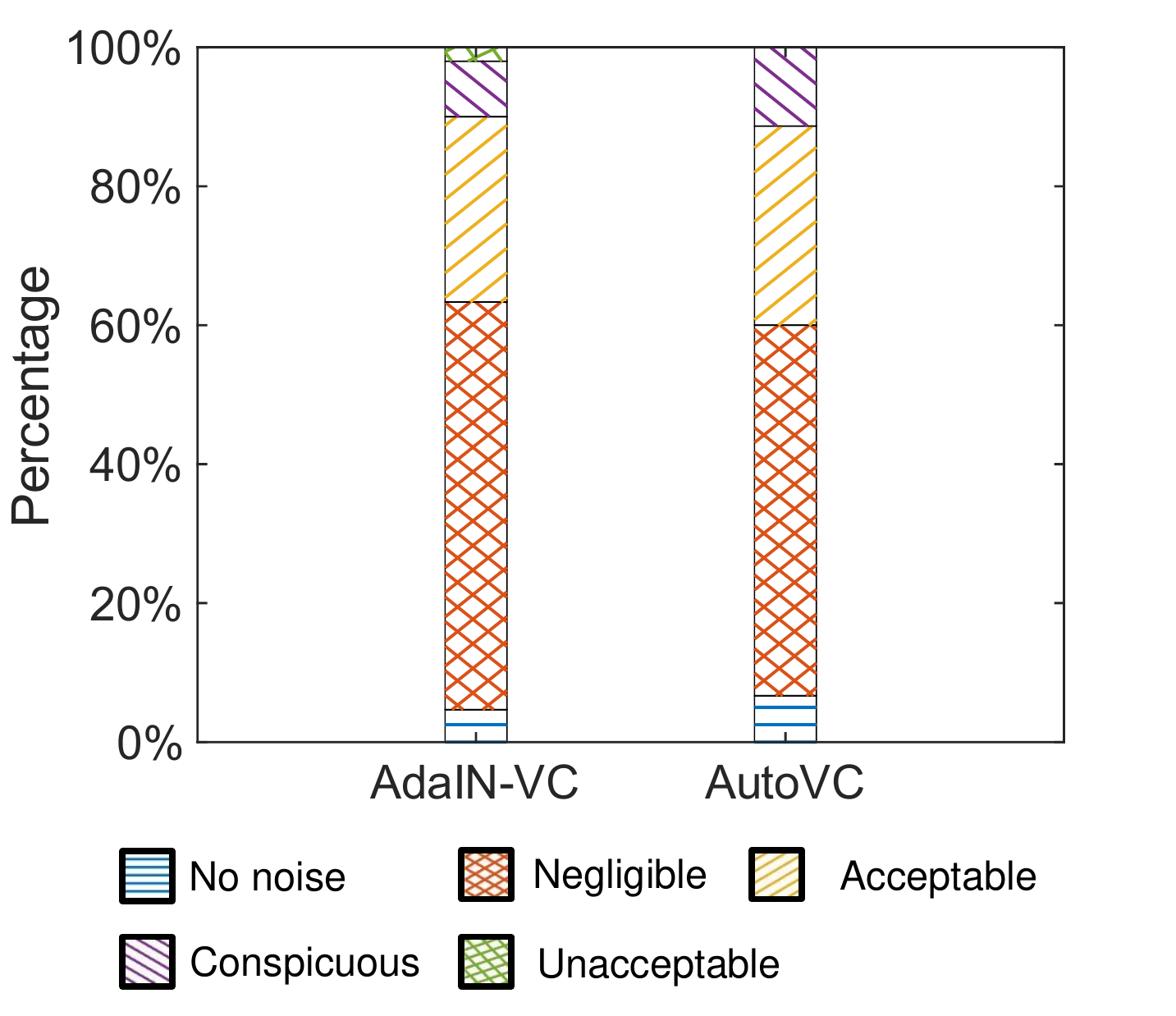}\label{fig:hs1}}\quad
    \subfigure[Human evaluation of \vm protection performance on AdaIN-VC model. ]{\includegraphics[height=4.5cm]{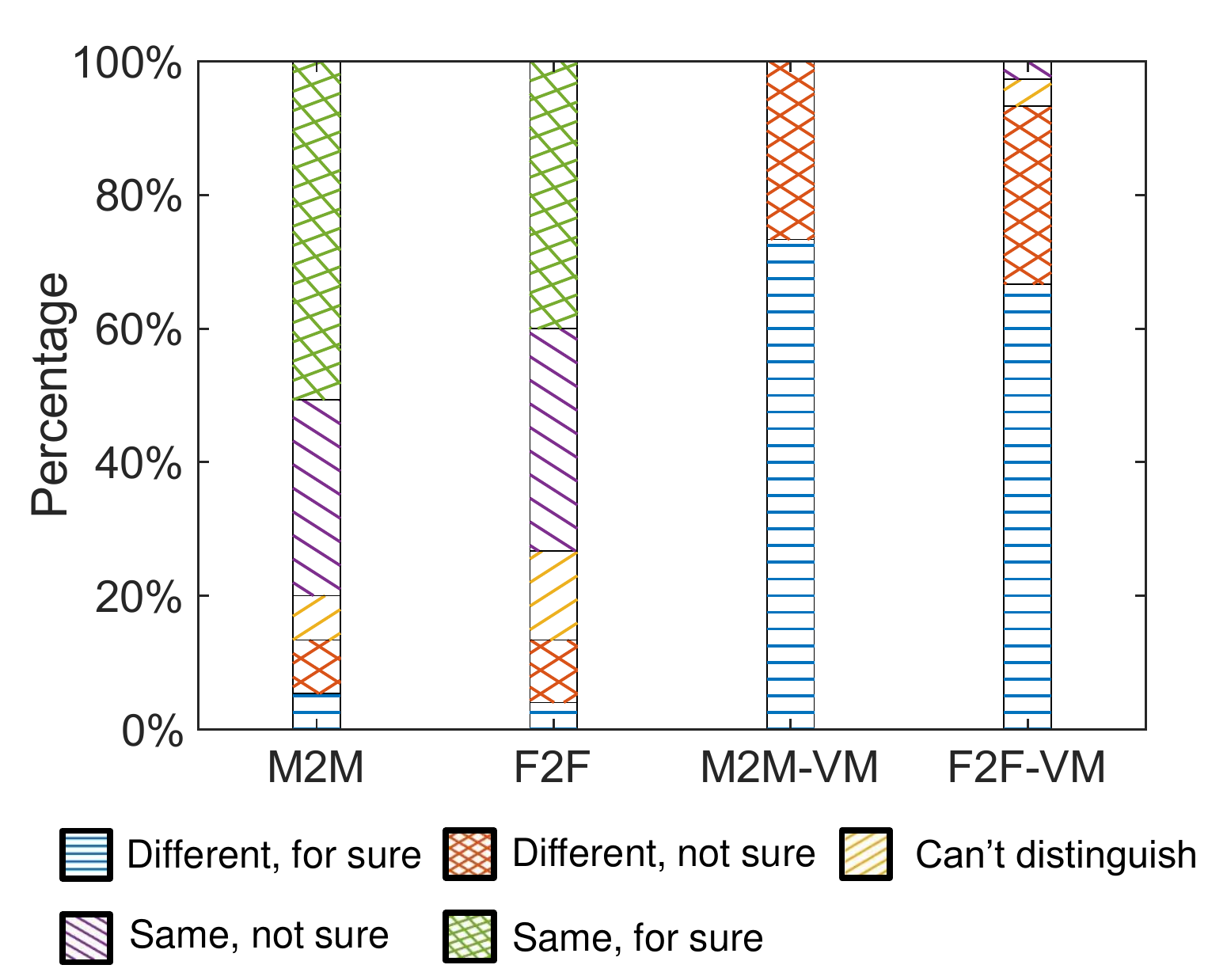}\label{fig:hs2}}\quad
    \subfigure[Human evaluation of \vm protection performance on AutoVC model. ]{\includegraphics[height=4.5cm]{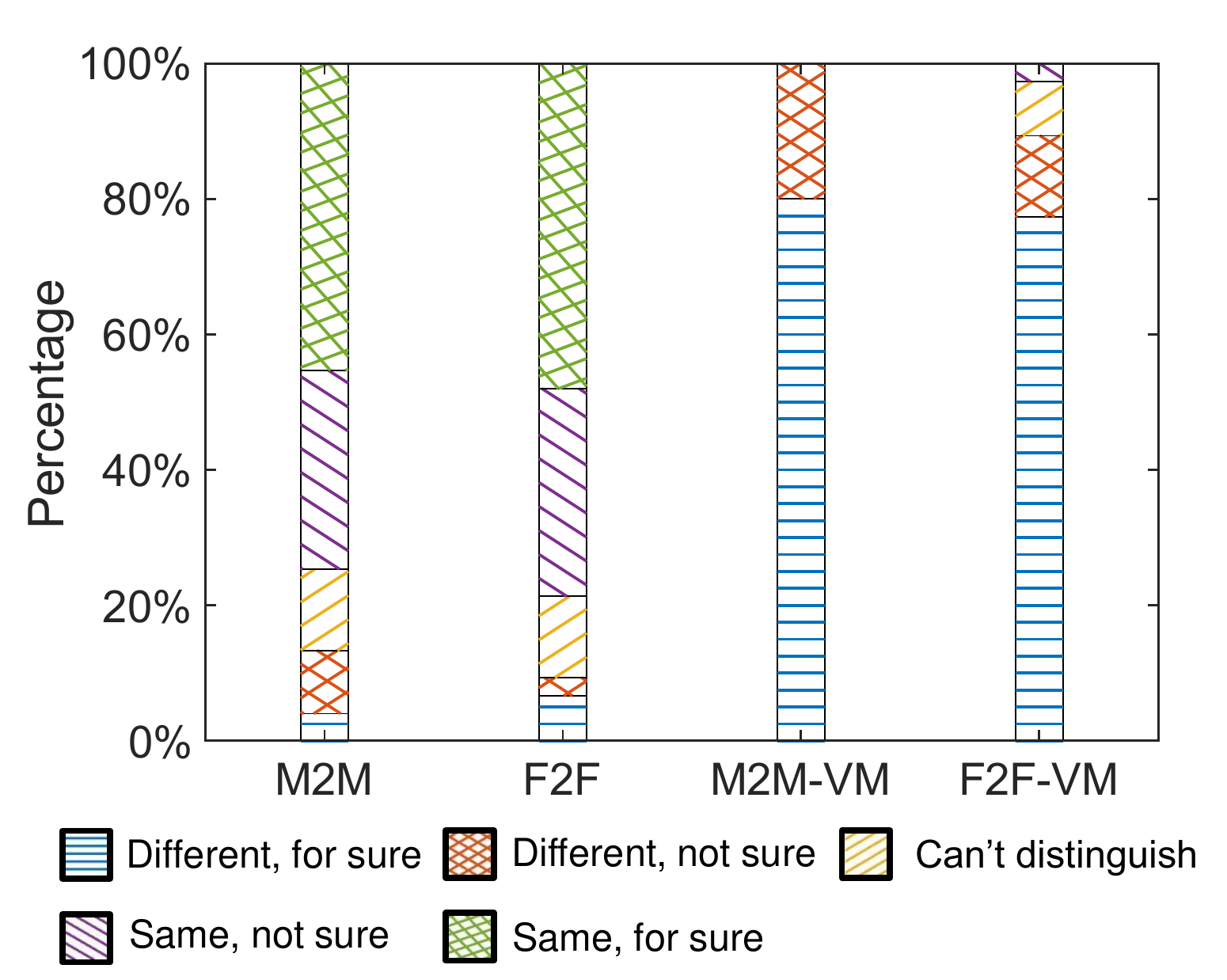}\label{fig:hs3}}
    \caption{Subjective human evaluation results about the protected speech noise level and the protection performance of \vm.}
    \label{fig: human study}
\end{figure*}

\vspace{-8pt}
Finally, we compare the offline PGD method performance with \vm. 
Theoretically, offline PGD method achieves the best performance for defending against voice synthesis attacks.
No synthetic audio from offline PGD protected speech can pass ASV systems.
Its performance can be further enhanced by increasing the iterations.
But for 1,500 iterations, its time consumption is approximately $20 \times$ audio length.
The computation time could be much longer when the model is deployed on devices without high performance GPU.
We mark \vm performance in \textbf{bold} in Table~\ref{Tab:overall performance}.
Compared with all other online baseline methods, \vm has the strongest protective impact.
And compared with offline PGD method, \vm achieves similar offline protection performance without incurring any time delay.
For both intra-gender and inter-gender voice conversions, no synthetic speech samples can pass ASV system.
Therefore, the results demonstrate that \vm can successfully protect human speech against unauthorized voice synthesis attacks.

\vspace{-6pt}

\subsection{\vm Evaluation on Different Voice Synthesis Models}
\label{section:different models}
In this section, we evaluate \vm performance on different voice synthesis models.
From Section~\ref{sec:vm_asv}, we learn that for voice conversion models, intra-gender voice conversion has better similarity than inter-gender voice conversion.
\rev{For AutoVC model we generate 50 intra-gender synthesized speech samples for each speaker.}

Fig.~\ref{fig:score} and Fig.~\ref{fig: rates} show the synthetic speech similarity and ASR before \vm protection.
For VC models, after being trained with data in VCTK corpus, both AdaIN-VC and AutoVC synthetic voice samples can pass ASV with more than $90\%$ success rate.
AutoVC shows smaller variance because of its well-designed bottle-neck mechanism.
But AdaIN-VC has better overall performance on similarity scores.
We also test TTS model and compare its performance with VC models.
The SV2TTS model is trained with 16 kHz speech samples.
\rev{To guarantee its best performance, we select 60 victim speakers (30 males and 30 females) from LibriSpeech dataset.
Also, we generate 50 synthesized speech samples from each speaker for testing.}
Then, we test all generated text-to-speech samples in ASV system.
Compared with VC models, regardless of the input phrase length, SV2TTS achieves the best similarity score and highest ASR.
The potential reason is that SV2TTS generates speech content directly from text input, so that it contains no prior existing voice pattern.
But for VC models, even though they are well optimized, there is residual voice pattern from the source speech $\textbf{p}$.
This will deteriorate the voice similarity of the synthetic speech and victim speaker.

In contrast, we also display the similarity score and ASR after \vm protection.
For VC models, \vm substantially reduces the similarity score between synthetic speech and the raw speech from the victim speaker.
None of the synthetic speech samples could compromise ASV system.
Since the synthetic voice patterns in SV2TTS are entirely determined by the reference speech $\textbf{x} + \delta$, this model is more susceptible to \vm protection compared to VC models.
Overall, under \vm protection, it is unlikely for attackers to spoof speaker verification system by either VC or TTS.
\vspace{-13pt}

\subsection{Human Study}
\vspace{-3pt}
Recent work~\cite{wenger2021hello} shows voice synthesis attack not only compromises ASV systems, but also deceives human ears.
In this section, we run a human study of \vm\footnote{The IRB request has been approved by the university board.}.
We select AdaIN-VC and AutoVC for synthetic voice generation because voice conversion models have more natural utterance than TTS.

In the human study evaluation, we recruit 25 anonymous volunteers with normal hearing ability (14 males and 11 females, ages from 20 to 38 years old), and ask them to listen to 12 groups of speech samples through the same loudspeaker with fixed volume ($\sim$60 dBA).
6 groups of synthetic speech samples are from AdaIN-VC model and 6 groups are from AutoVC model.
For each model, half of samples are male-to-male and the other half is female-to-female.
Each group contains 4 audio samples: \ding{192} raw speech audio without protection, \ding{193} the same speech audio but with \vm protection, \ding{194} synthetic voice sample from the raw speech, \ding{195} the same synthetic voice sample from \vm protected speech.
All the speech sample lengths are from 4 seconds to 8 seconds.

First, we ask the volunteers about the noise level in \vm protected speech. After listening to the raw speech and protected speech, the listeners are suppose to measure the perturbation perceptibility of the protected sample.
Options from low to high are: (A) Cannot hear any noise;
(B) Can hear noise, but negligible;
(C) Can hear noise, but acceptable;
(D) Can hear conspicuous noise;
(E) Very strong noise, unacceptable.

We collect all answers and present them in Fig.~\ref{fig:hs1}.
The results depict that for both AdaIN-VC and AutoVC, the noise level in the protected voice samples is acceptable for human ears.
Around 60\% answers are "negligible" or "imperceptible", which means that \vm protected speech audio has similar clarity as raw audios.
More than 90\% answers consider the noise in protected audios is acceptable for human ears. 
Only 2 out of 300 answers point out that the protected speech has "unacceptable" noise.

Next, we ask the volunteers to evaluate the similarities between the raw speech and synthetic speech samples.
Similarities options from low to high are: (A) Different voice, for sure;
(B) Different voice, but not sure;
(C) Cannot distinguish;
(D) Same voice, but not sure;
(E) Same voice, for sure.
Consider the voice conversion performance difference caused by different speaker genders, we separately display results of male-to-male and female-to-female speech samples.

Fig~\ref{fig:hs2} and Fig~\ref{fig:hs3} show the voice similarity between the \ding{192} raw audio  \& \ding{194} synthetic audio from raw speech, and \ding{192} raw audio \& \ding{195} synthetic audio from protected speech.
For both AdaIN-VC and AutoVC, most synthetic speeches from raw speech can spoof human ears.
More than 75\% answers indicate that the synthetic speech is definitely or probably from the same speaker as the raw audio.
Only 4\% answers indicate that the voice is absolutely from another speaker.
Therefore, the results show that synthetic voice has sufficient similarity as the victim speaker when launching intra-gender voice conversion.
In contrast, for synthetic speech samples from \vm protected speech (M2M-VM and F2F-VM), they could not fool human ears any more.
More than 75\% answers believe that the synthetic audios from \vm protected speeches have definitely different voice from the victim speaker.
In total, over 97\% answers claim that they will doubt the speaker identity behind the voice.
Nobody consider that the synthetic speech matches the victim's voice very well.
\newrev{Furthermore, the offline PGD method has been shown to achieve comparable protection performance in preventing deepfake voices from deceiving human ears~\cite{huang2021defending}.
In addition, we present the impact of weight-based noise mitigation in Appendix~\ref{weight_appendix}.}

\vspace{-12pt}
\subsection{Cross-model Evaluation}

Attack-VC~\cite{huang2021defending} evaluates defense performance under black-box scenarios by training substitute models.
However, sometimes we have no information about which voice synthesis model will be implemented.
In this section, we evaluate \vm performance under cross-model conditions.

We apply \vm to generate adversarial speech samples targeting one voice synthesis model, and apply them on other models.
The cross-model ASR is listed in Table~\ref{Tab:crossmodel}.
Overall, \vm is still effective for cross-model scenario.
But the performance is degraded as the result of different encoder models, preprocessing methods, and training data.
For example, AdaIN-VC applies 512-dim mel-spectrogram as speaker encoder input, but 80-dim for AutoVC speaker encoder.
Also, SV2TTS is trained with 16 kHz audios, but AdaIN-VC and AutoVC use 48 kHz audios.
When we input 16 kHz speech sample in AutoVC, it could not output qualified audios.
Moreover, the adversarial samples targeting AdaIN-VC can reduce the similarity of AutoVC synthetic speech as well.
When we use adversarial samples from AdaIN-VC or AutoVC on SV2TTS model, a few synthetic samples can still compromise ASV system.
\rev{We can further apply AdaBelief method~\cite{zhuang2020adabelief} to improve the transferability of protected samples by gradually reducing the learning rate, which we leave for future work.}

\begin{table}[t]
    \setlength{\abovecaptionskip}{0pt} 
    \setlength{\belowcaptionskip}{0pt}
\centering
\caption{We generate the adversarial examples from the source model and apply the target model for voice synthesis. The result shows that \vm can still reduce the ASR under cross-model setup.}
\label{Tab:crossmodel}
\small
\begin{tabular}{c|ccc} 
\toprule
\diagbox{Target}{Source} & AdaIN-VC     & AutoVC       & SV2TTS        \\ 
\midrule
AdaIN-VC                 & - & 15.0\%       &  10.5\%            \\ 
\midrule
AutoVC                   & 12.8\%       & - & 0.0\%         \\ 
\midrule
SV2TTS                   & 7.3\%       & 15.2\%            & - \\
\bottomrule
\end{tabular}
\end{table}
\subsection{Adaptive Attack Evaluation}
Denoisers are widely deployed in speech enhancement models for audio quality improvement.
Adversarial perturbations can also be mitigated by some audio signal transformation methods.
To evaluate \vm robustness against adaptive attackers, we apply denoiser and WaveGuard~\cite{272238} to remove the perturbation in protected audios and launch voice synthesis by AdaIN-VC model. 
All synthetic speech samples are from intra-gender speech synthesis.

A simple denoiser fails to remove the well-designed perturbation in protected speech, and the synthetic speech similarity is even further degraded since the denoiser eliminates some low-power signals in the speech.
Typically, speech-to-text models can still accurately transcribe speech with lower resolution. This allows WaveGuard to compromise speech audio quality while mitigating the adversarial perturbation.
\rev{
We evaluated four WaveGuard signal transformation methods, including Downsampling-Upsampling with different frequencies, Quantization-Dequantization, Mel-spectrogram conversion with different mel bins, and Linear Predictive Coding (LPC) with different orders, on speech samples protected by \vm. 
We list the best similarity score in Table~\ref{Tab: waveguard}.
In contrast to adversarial examples that target speech-to-text models, WaveGuard methods cannot undermine the protection performance of \vm since \vm optimizes the perturbation in mel-spectrogram, which is resilient to regular signal transformation techniques.
In this way, the compression or transformation of audio cannot accurately restore the initial clear speech. 
Furthermore, WaveGuard reduces the audio quality of the synthetic speech. 
As a result, \vm maintains a high level of robustness against denoising and signal transformation approaches.
For sophisticated attackers, it is possible to recover the raw audio by reverse engineering if they have full knowledge about \vm.
Other approaches, for example, randomized smoothing~\cite{cohen2019certified}, adversarial training~\cite{madry2018towards} and diffusion model~\cite{wu2023defending} are potential solutions to evade \vm.}

\begin{table}
\centering
\footnotesize
\caption{\vm performance when adaptive attackers apply denoiser and WaveGuard.}
\label{Tab: waveguard}
\begin{tabular}{ccccccc} 
\toprule
\multirow{2}{*}{\begin{tabular}[c]{@{}c@{}}~Adaptive \\methods~\end{tabular}} & \multirow{2}{*}{None} & \multirow{2}{*}{Denoiser} & \multicolumn{4}{c}{WaveGuard}                                                                                                                                                                                                             \\ 
\cmidrule{4-7}
                                                                              &                       &                           & \begin{tabular}[c]{@{}c@{}}Down-Up\\ ($f$=24k)\end{tabular} & \begin{tabular}[c]{@{}c@{}}Quan.-\\Dequan.\end{tabular} & \begin{tabular}[c]{@{}c@{}}Mel.\\(Bin=128)\end{tabular} & \begin{tabular}[c]{@{}c@{}}LPC\\(Ord.=10)\end{tabular}  \\ 
\midrule
Score                                                                         & 0.096                 & 0.090                     & 0.078                                                   & 0.082                                                   & 0.080                                                     & 0.073                                                     \\
\bottomrule
\end{tabular}
\end{table}
\vspace{-5pt}

%% file: document/5_related_work.tex
\section{Discussion and Limitations}
\label{discussion}
\begin{figure}
    \setlength{\abovecaptionskip}{0pt} 
    \setlength{\belowcaptionskip}{-20pt} 
    \centering
    \subfigure[Physical world experiment setup.]{\includegraphics[height=2.8cm, width = 4cm]{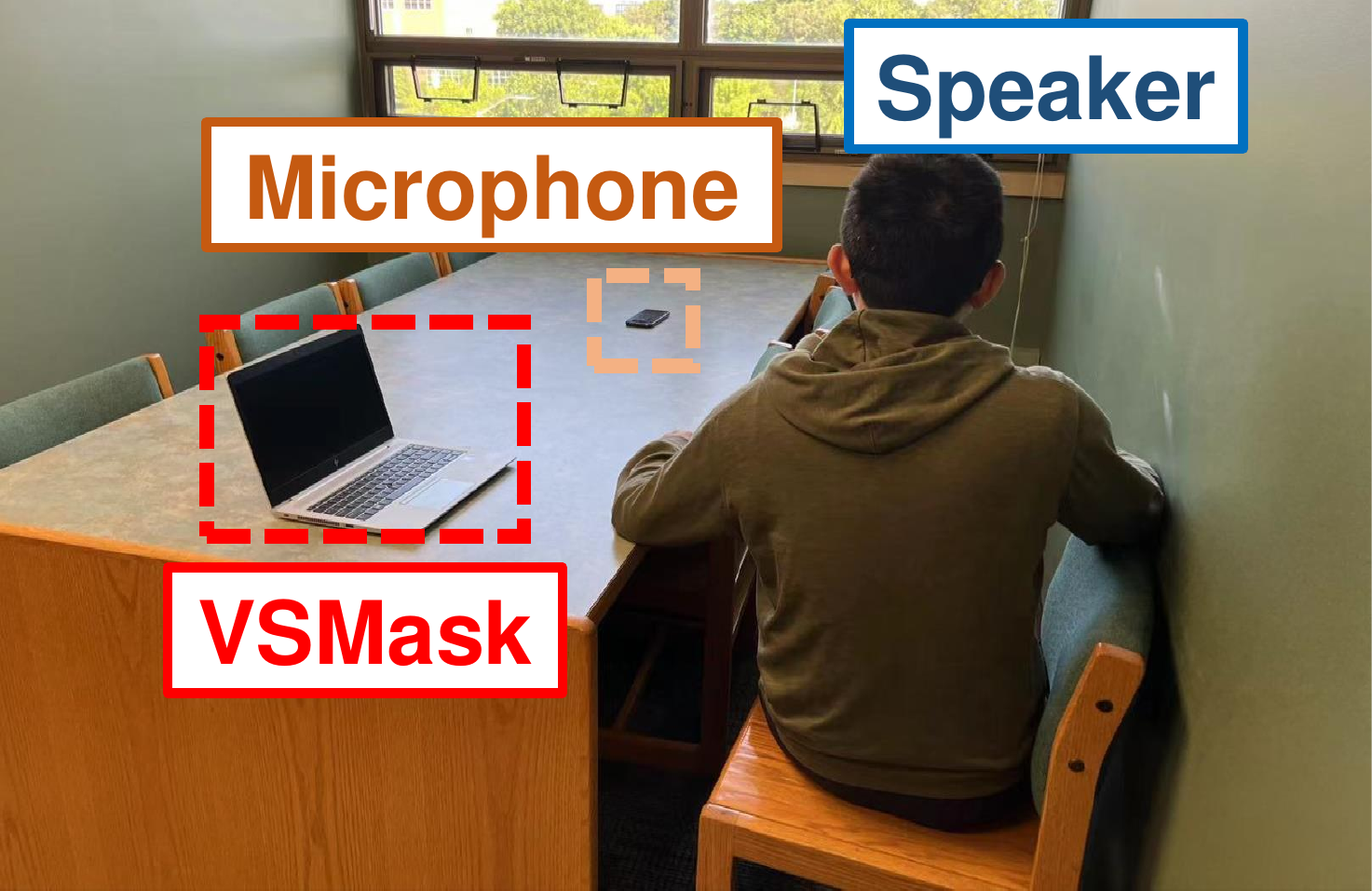}\label{fig: real scenario}}
    \quad
    \subfigure[\vm protection performance in real-world scenario.]
     {\includegraphics[height=3cm]{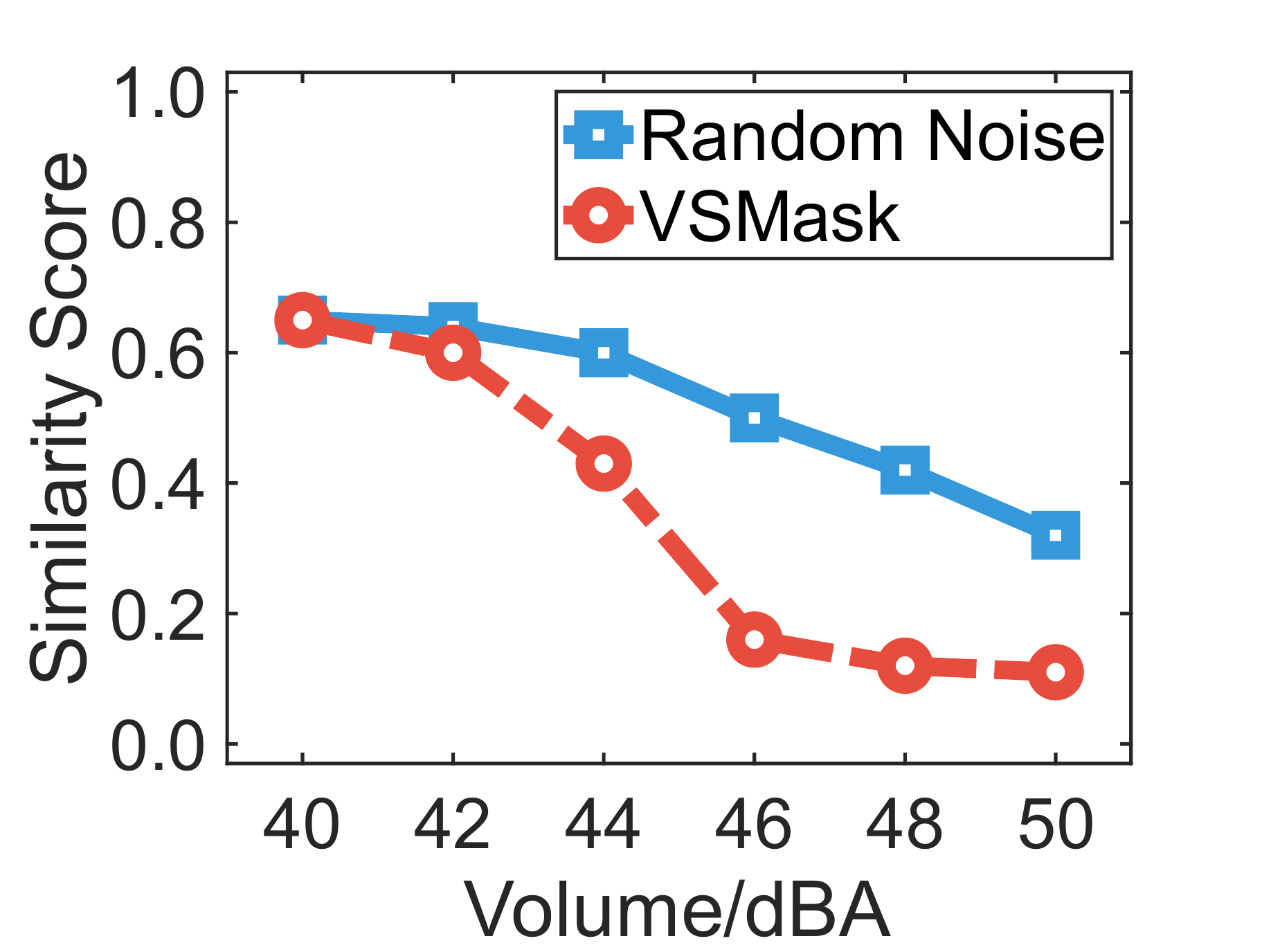}\label{fig: real-world-scores}}
    \caption{We implement \vm in real-world scenario, where it shows much stronger protection impact than white noise.}
    \label{Fig: real-world all}
\end{figure}

\noindent \textbf{Defense in Real-world Scenarios}.
Fig.~\ref{fig: real scenario} displays the setup of physical world \vm protection.
The speaker is reading transcripts from VCTK dataset in a quiet room.
Meanwhile, \vm is deployed on a laptop 50 cm away from the speaker.
At the same time, an eavesdropper is stealthily recording the speech audio through an iPhone 13 one meter away from the speaker and \vm.
Then, the eavesdropper will synthesize the speaker's voice in the recorded speech by AdaIN-VC model.
Meanwhile, we put a sound level meter close to the microphone to measure the volume of sounds.
The background noise in the room is \emph{40} dBA, and we maintain the speech loudness around \emph{75} dBA.
While the victim is talking in quiet room, the attacker can eavesdrop clear speech signal and generate qualified deepfake voice.

Notably, in real-world scenarios, different from cyberspace, the perturbation $\delta$ is usually distorted during over-the-air transmission.
\rev{To overcome the challenge, we implement a band pass filter (500 $\sim$ 4000 Hz) to filter vulnerable frequencies, and room impulse response (RIR) filter to mitigate the signal distortion in airborne transmission~\cite{yakura2018robust}.
When \vm is applied in a new environment, the RIR filter should be redesigned.}
Fig.~\ref{fig: real-world-scores} shows the protection impact under real-world scenarios.
When the average perturbation loudness of \vm reaches 46 dBA, it can effectively compromise the target voice synthesis model without affecting the speech intelligibility.
In comparison, random white noise with the same volume fails to degrade the similarity score much.
When the perturbation volume reaches 50 dBA, \vm can successfully reduce the ASR to 0\% while most synthetic speech generated from white noise masked speech can still bypass the ASV model.
The experiment is launched in a quiet environment.
If the environment is noisy (Volume > 50 dBA), the synthesized speech quality will also be affected even without \vm.
\rev{In particular, adaptive attackers may use a microphone array to denoise \vm perturbation by phase cancellation~\cite{599604}.
We can enhance the spatial complexity and increase the difficulty of \vm perturbation cancellation by utilizing multi-channel loudspeakers.}

\rev{\noindent \textbf{Real-time Feasibility}.
Training a predictive model for one speaker on an NVIDIA A6000 GPU costs less than 5 minutes. In the inference process, one single computation of the predictive model costs $\sim$ 13 milliseconds on a common CPU. 
We also measure the time delay caused by bidirectional transformation between audio signal and mel-spectrogram, which is $\sim$ 200 milliseconds. 
Therefore, the total time cost for generating voice perturbation is around 0.2 seconds, and the latency of 0.4 seconds is sufficient for \vm to achieve real-time applicability.}

\noindent \textbf{Adversarial Training}.
Adversarial training can effectively improve the robustness of deep learning models~\cite{madry2018towards}.
Adaptive attackers can apply adversarial training to improve the similarity between the synthetic voice and the victim speaker protected by \vm.
However, adversarial training inevitably reduce the synthesis performance when dealing with clear speech audios.
Moreover, recent work~\cite{zheng2021black} shows that adversarial training takes up to 10 days with a powerful GPU group even for a mini dataset.
Therefore, given \vm's real-time protection feature, adversarial training is not a realistic solution that can be adopted by attackers to defeat \vm.

\rev{\noindent \textbf{Limitations}.
\vm has several limitations. 
First, specially designed adaptive attacks are not considered in this paper, e.g., adversaries who have full knowledge about \vm. The adaptive attackers can purify the speech by a well-designed neural network or diffusion models, and we leave the exploration of these evasive approaches in future work.
Second, we only launch real-world evaluation in a single room in the physical experiments. In different environments, the performance of \vm might be different.
Third, \vm cannot provide 100\% successful protection regardless of the voice synthesis model.
A few synthetic speech can still pass ASV systems under pure black-box setup.
Finally, the proposed method does not provide a formal privacy guarantee for human speech. This paper aims to conduct empirical privacy analysis with \vm and provide a practical solution to enhance speech privacy in real-time physical and digital scenarios.}

\vspace{-9pt}
\section{Related Work}
\vspace{-3pt}
\subsection{Voice Synthesis}
For traditional VC methods, it is necessary to train the model with parallel speech data from the source speaker and victim speaker.
Chen et al.~\cite{chen2003voice} propose a voice conversion model based on Gaussian Mixture Models (GMM).
 Erro et al.~\cite{erro2009voice} apply a weighted frequency wrapping method to change in voice in the speech.
But these methods suffer from over-smoothing~\cite{helander2010voice} and unnatural voice.
It is also difficult to collect the parallel speech data from different speakers~\cite{mohammadi2014voice, takamichi2014postfilter}.

To address such challenges, Variational Autoencoder (VAE) is introduced to convert human voice in the speech~\cite{hsu2016voice}.
Compared with traditional voice conversion models, VAE is able to learn independent speech content regardless of the speaker identity.
Therefore, VAE shows better conversion efficiency and requires no parallel speech data for training.
Hsu et al.~\cite{hsu2017voice} propose a voice conversion model that generates converted voice frame-by-frame.
AdaIN-VC~\cite{chou2019one} significantly improves the converted speech quality by adding an extra adversarial training stage.
Cycle-GAN~\cite{kaneko2017parallel} and Star-GAN~\cite{kameoka2018stargan} also apply adversarial learning method in VAE based voice conversion. 
One of the state-of-the-art work is AutoVC~\cite{qian2019autovc}, which is a voice conversion model that introduces bottleneck mechanism into VAE.
It successfully achieves zero-shot conversion and balances the speaker disentanglement and voice conversion quality via well-designed bottleneck.
Moreover, generative adversarial network (GAN) itself can be applied to clone human voice~\cite{gao2018voice}.
But it achieves limited conversion quality.
TTS can also synthesize speech with real human voice.
Wavenet~\cite{oord2016wavenet} introduces a generative model to output raw audio according to the input text.
Tacotron 2~\cite{shen2018natural} further inverts Wavenet generated spectrograms with attention.
But Tacotron 2 can only generate speech with one default speaker voice.
Deep Voice 2~\cite{gibiansky2017deep} is a speech generative model that supports multiple speaker, while Deep Voice 3~\cite{ping2017deep} uses a encoder-decoder based neural network.
It can apply TTS synthesis with over 2,400 different voices from LibriSpeech.
SV2TTS~\cite{jia2018transfer} introduces a TTS model that enables speech synthesis for an arbitrary speaker.
It shares similar speaker encoder module as voice conversion models.
Wenger et al.~\cite{wenger2021hello} comprehensively evaluate the voice synthesis threat in the real world scenarios.
\vspace{-10pt}
\subsection{Adversarial Speech Attacks and Defenses}
Existing work demonstrates that neural networks can be easily fooled by adversarial examples.
Carlini et al.~\cite{carlini2018audio} point out that adversarial speech samples can compromise ASR systems.
Sch{\"o}nherr et al.~\cite{schonherr2018adversarial} demonstrate that ASR can be compromised by perturbation out of human hearing range.
Adversarial speech samples can be hidden in speech~\cite{197215} or music~\cite{yuan2018commandersong}, and deliberately crafted to deceive speech recognition systems. 
In some cases, these adversarial samples can be made imperceptible to the human ear~\cite{qin2019imperceptible}, posing security threats in both digital and physical environments. 

Similarly, speaker recognition models are also vulnerable to adversarial attacks~\cite{kreuk2018fooling,gong2017crafting}.
Chen et al.~\cite{chen2021real} proposes FAKEBOB, an adversarial attack successfully compromise speaker recognition systems under black-box and over-the-air attack scenarios.
Zhang et al.~\cite{zhang2020voiceprint} also successfully spoof commercial speaker verification systems by well-designed adversarial speech examples.

Offline adversarial speech attacks only work when the entire speech is available.
To launch real-time attacks toward streaming speech signal, Lu et al.~\cite{lu2021exploring} introduce universal adversarial perturbation, which could force the ASR models to output the target transcript regardless of the real input is.
AdvPulse~\cite{li2020advpulse} successfully fools ASR system by a short adversarial ``pulse".
The adversarial perturbations successfully mislead the ASR system without signal synchronization.
But these universal perturbations work only if the crafted output is short.
Gong et al.~\cite{gong2019real} introduce a real-time attack targeting streaming voice signal.
This real-time attacks works well for keyword detection applications with a pre-defined dictionary.
SpecPatch~\cite{guo2022specpatch} proposes an adversarial attack could mute the upcoming voice commands and compromise long speech samples.
VoiceCamo~\cite{chiquier2021real} proposes a real-time attack based on a predictive model.
It leverages the past speech signal as input and predict the most effective perturbation to generate adversarial example along with the upcoming speech.
\vspace{-10pt}
\subsection{Speech Privacy Enhancement}
\rev{Qian et al.~\cite{qian2018hidebehind} leverage voice conversion to hide speaker's identity.
McAdams~\cite{patino2020speaker} and Vaidya et al.~\cite{vaidya2019you} mask human speech by signal processing to enhance speech privacy.
Han et al.~\cite{10.1145/3372297.3420025} propose a differential privacy based fast speech de-identification system to provide formal privacy guarantee on an untrusted server.
Emotionless~\cite{aloufi2019emotionless} is a privacy-preserving speech analysis tool for voice assistants  based on differential privacy. In contrast, this paper aims to develop a practical speech protection system to offer privacy enhancement for real-world speech applications. }

Meanwhile, adversarial examples can be leveraged to enhance speech privacy.
V-Cloak~\cite{deng2022v} presents a timbre-preserving model to achieve voice anonymization, but its protection performance against voice synthesis attacks is not evaluated.
Attack-VC~\cite{huang2021defending} introduces a defense method based on adversarial examples, which could protect offline human speech against voice synthesis attacks. 
\vm introduces a novel combination of universal perturbation and predictive models to protect the real-time speech from voice synthesis attacks.

%% file: document/6_conclusion.tex
\vspace{-10pt}
\section{Conclusion}
Voice synthesis attack is a major threat to the voice controllable systems. 
This research aims to prevent the attackers from synthesizing the protected voices. 
Specifically, we explore the feasibility of real-time defense against voice synthesis.
By implementing real-time predictive model and universal perturbation header, \vm successfully protects real-time speech signal in both online and offline real-world scenarios.
Moreover, \vm mitigates the audio distortion in protected speech via the weight-based amplitude constrained mechanism.
We evaluate \vm's voice protection performance under different real-world scenarios, and the experimental results show that \vm can effectively defend against voice synthesis attacks in live speech scenarios.
\vspace{-10pt}
\section*{Acknowledgement}
We would like to thank the anonymous reviewers and shepherd for their insightful comments and feedback on our work. This work was supported in part through National Science Foundation grants CNS-1950171 and CISCO research award.